\documentclass[12pt]{article}
\usepackage{color,amssymb,epsfig,verbatim,amsmath}
\usepackage{amsfonts} % used for "bold fATE in greek words"
\usepackage{longtable}
\usepackage{booktabs}
\usepackage{adjustbox}
\usepackage{accents}
\usepackage{pdflscape}
\usepackage{tabularray}
\usepackage{accents}
\usepackage{array}
\usepackage{pdfpages}
\usepackage{tabu}
\usepackage[title]{appendix}
\newcommand{\vecTheta}{\mbox{vec}\left( \boldsymbol{\Theta} \right)}
\newcommand{\vecThetah}{\mbox{vec}( \widehat{\boldsymbol{\Theta}} )}

\def\dfrac#1#2{{\displaystyle{#1\over#2}}}

\def\boxit#1{\vbox{\hrule\hbox{\vrule\kern6pt
          \vbox{\kern6pt#1\kern6pt}\kern6pt\vrule}\hrule}}
\def\refhg{\hangindent=20pt\hangafter=1}
\def\refmark{\par\vskip 2mm\noindent\refhg}

\def\refhg{\hangindent=20pt\hangafter=1}
\def\refmark{\par\vskip 2mm\noindent\refhg}

\def\bse{\begin{eqnarray*}}
\def\ese{\end{eqnarray*}}
\def\be{\begin{eqnarray}}
\def\ee{\end{eqnarray}}
\def\bq{\begin{equation}}
\def\eq{\end{equation}}
\def\bse{\begin{eqnarray*}}
\def\ese{\end{eqnarray*}}

\newtheorem{Exa}{Example}

\usepackage{pgf,tikz}
\usetikzlibrary{arrows}
\usepackage{hyperref}
\hypersetup{
    colorlinks=true,
    linkcolor=blue,
    filecolor=magenta,      
    urlcolor=blue,
}
\usepackage{lscape}
\usepackage{pdflscape}
\usepackage{afterpage}

\newtheorem{theorem}{Theorem}[section]
\DeclareMathOperator*{\argmin}{argmin}
\DeclareMathOperator*{\argmax}{argmax}

\makeatletter
\newcommand*\bigcdot{\mathpalette\bigcdot@{.5}}
\newcommand*\bigcdot@[2]{\mathbin{\vcenter{\hbox{\scalebox{#2}{$\m@th#1\bullet$}}}}}
\makeatother

\begin{document}
\thispagestyle{empty}

\hfill March 27, 2025 \\ \\

\baselineskip=28pt
\begin{center}
{\LARGE{\bf A Unified Approach for Estimating Various Treatment Effects in Causal Inference}}
\end{center}
\baselineskip=14pt
\vskip 2mm
\begin{center}
Kuan-Hsun Wu and Li-Pang Chen\footnote{Correpsonding Author. Email: \hyperlink{mailto:lchen723@nccu.edu.tw}{lchen723@nccu.edu.tw}}
\\~\\

\textit{Department of Statistics, National Chengchi University}
\end{center}
\bigskip

\begin{center}
{\Large{\bf Abstract }}
\end{center}
\baselineskip=17pt
{
In this paper, we introduce a unified estimator to analyze various treatment effects in causal inference, including but not limited to the average treatment effect (ATE) and the quantile treatment effect (QTE). The proposed estimator is developed under the statistical functional and cumulative distribution function structure, which leads to a flexible and robust estimator and covers some frequent treatment effects. In addition, our approach also takes variable selection into account, so that informative and network structure in confounders can be identified and be implemented in our estimation procedure. The theoretical properties, including variable selection consistency and asymptotic normality of the statistical functional estimator, are established. Various treatment effects estimations are also conducted in numerical studies, and the results reveal that the proposed estimator generally outperforms the existing methods and is more efficient than its competitors.}

\par\vfill\noindent

\par\medskip\noindent
\underline{\bf Short title}: Estimating Various Cauasl Effects Using IPW CDF

\clearpage\pagebreak\newpage
\pagenumbering{arabic}

\newlength{\gnat}
\setlength{\gnat}{22pt}
\baselineskip=\gnat

\clearpage

\section{Introduction} 
{

Causal inference aims to explore the causality of two variables, and it attract people's attention in various fields recently, such as social science or biological studies. One of the important questions in causal inference us the estimations of treatment effects, which reflect how different treatments affect the outcomes. In the literature, some typical causal effects include but are not limited to the average treatment effect (ATE, Chen, 2020; Yi and Chen, 2023) and the quantile treatment effect (QTE, Firpo 2007; Donald and Hsu, 2014; Hsu et al., 2022). To estimate treatment effects, the inverse probability weight (IPW) method (e.g., Rosenbaum and Rubin, 1983) is one of the popular approaches, whose key idea is to model the propensity score and treat it as the weight to recover the "missingness" in the outcome. This approach has also been widely used to the estimation procedure, such as ATE (e.g., Chen, 2020; Yi and Chen, 2023), QTE (e.g., Firpo 2007; Donald and Hsu, 2014; Hsu et al., 2022) and treatment effect for survival data (e.g., Chapfuwa et al., 2021).

In applications, one may encounter a scenario that the dataset contains multivariate or high-dimensional confounders. Among those confounders, few of them are informative to the outcome, so it is crucial to do variable selection. There are some methods available to address this challenge.   To name a few, Ertefaie et al. (2018) proposed the weighted LASSO penalization with simultaneous consideration on outcome and treatment models. Yi and Chen (2023) proposed the penalized likelihood function with error-prone confounders. Following the spirit of adaptive LASSO from Zou (2006), the outcome-adaptive LASSO proposed by Shortreed and Ertefaie (2017) focused on the estimation of propensity score with the logit model adopted.  Moreover, Bayesian approaches were also employed to causal inference, including Koch et al. (2020) who implemented variable selection under the similar framework of estimating causal effect with spike and slab priors applied.

In addition to selecting informative covariates, the other challenging feature in multivariate or high-dimensional variables is the complex network structure. To offer an intelligible interpretation of the dependencies among high-dimensional confounders,  \textit{graphical model} is a powerful tool and has been adopted for numerous applications in regression models or supervised learning; see details in Chen (2024, Section 4). However, the impact of network structure in confounders and the resulting estimation for treatment effects have not been fully explored. 

To address the challenges in the multivariate or high-dimensional settings and unify various treatment effect estimation into a general estimator, we proposed the IPW method under the cumulative distribution function (CDF) and statistical functional form. The proposed estimator is robust in dealing with outlier and is reduced to some well known treatment effects if the statistical functional is properly specified. Moreover, we also implement variable selection technique to detect informative confounders and network structure when estimating the propensity score. Theoretically, we establish the variable selection consistency and asymptotic properties for the proposed estimator.

The remainder is organized as follows. In Section~\ref{Sec2}, we introduce the fundamentals of causal inference including necessary assumptions and estimands of interest, along with the model formulation of network structure. Subsequently, we introduce the proposed approach and the theoretical results in Section~\ref{Sec3}. To verify the effectiveness of our method, we conduct the simulation studies and real data analysis in Sections~\ref{Sec4} and~\ref{CHILD}, respectively. Finally, the highlights of this article are discussed and summarized in Section~\ref{Summary}.

\section{Preliminaries and Model Setup}\label{Sec2}

\subsection{Causal Effects and Propensity Score}\label{Sec2.1}
Let $\mathbf{X} =\left(X_1, \cdots, X_p \right)^\top$ denote the $p$-dimensional vector of covariates, or confounders with little case letter $\boldsymbol{x}$ being a realized value, and let $A \in \{0,1\}$ denote the binary treatment that indicates treatment ($A=1$) or control ($A=0$). Given the treatment effect $A = a$, let $Y^{(a)}$ denote the potential outcome. Specifically, when $a=1$, $Y^{(1)}$ stands for the potential outcome if a subject is assigned to the treatment; when $a=0$, then $Y^{(0)}$ is the potential outcome if a subject is assigned to the control. Moreover, let $F^{(a)}(y)$ denote the CDF of $Y^{(a)}$ for $a=0,1$. 

Let $T(F)$ denote the statistical functional, which is a function of the CDF $F(y)$. Hence, we consider a general class of the treatment effect
\begin{equation}\label{general treatment effect}
    T(F^{(1)}) - T(F^{(0)}).
\end{equation}
Then (\ref{general treatment effect}) can be reduced to some commonly used treatment effects in causal inference. For example, when $T(F) = \int y dF(y)$, then (\ref{general treatment effect}) is reduced to the {\it average treatment effect} (ATE), which is given by
\begin{eqnarray}\label{ATE}
\begin{aligned}
    \tau &\triangleq\int_{\mathbb{R}} y dF^{(1)}(y) - \int_{\mathbb{R}} y dF^{(0)}(y) \\
    &=E\left( Y^{(1)} \right) - E\left( Y^{(0)} \right).
\end{aligned}
\end{eqnarray}
When $T(F^{(1)})$ and $T(F^{(0)})$ are specified as the $q$th quantiles of $Y^{(1)}$ and $Y^{(0)}$, i.e.,
\begin{eqnarray}\label{quantile1}
    \xi^{(1)}(q)\triangleq \inf\bigg\{y\in \mathbb{R}\mid F^{(1)}(y)\geq q\bigg\}\text{ and }\xi^{(0)}(q) \triangleq \inf\bigg\{y\in \mathbb{R}\mid F^{(0)}(y)\geq q\bigg\}
\end{eqnarray}
for $q\in(0,1)$, respectively, the (\ref{general treatment effect}) becomes the $q$\textit{th quantile treatment effect} (QTE):
\begin{eqnarray}\label{QTE}
    \Xi(q) \triangleq  \xi^{(1)}(q) -  \xi^{(0)}(q).
\end{eqnarray}
If $T(F)$ is {the probability functional}, say $T(F)=\int_{-\infty}^ydF(x)$ for a fixed $y$, 
then (\ref{general treatment effect}) gives the distributional treatment effect (DTE), which is defined as 
\begin{eqnarray}\label{DTE}
   \Delta(y) \triangleq F^{(1)}(y) - F^{(0)}(y)
\end{eqnarray}
for a given $y\in \mathbb{R}$. 
\par
To estimate (\ref{ATE}), (\ref{QTE}) and (\ref{DTE}), a crucial issue is the estimation of $F^{(a)}(y)$. In this study, we employ the inverse probability weight (IPW) method. Specifically, the propensity score (PS) is defined as the conditional probability of treatment, given confounders, which is given by 
\begin{eqnarray} \label{PS}
\pi(\boldsymbol{x}) \triangleq P \left(A = 1 | \mathbf{X} = \boldsymbol{x} \right).
\end{eqnarray}
To further discuss the estimation as well as the relevant development, we impose the following conditions that are also commonly assumed in the framework of causal inference (e.g. Yi and Chen 2023):
\begin{itemize}
\item[(C1)] {\it Stable Unit Treatment Value Assumption (SUTVA)}: 

Response of a subject is not affected by responses of other subjects (noninterference). The treatment $A$ could be assigned by different ways, but they all lead to the same outcome (consistency). In this assumption, we have $Y = A Y^{(1)} + \left(1-A\right) Y^{(0)}$.
\item[(C2)] {\it Strong Ignorable Treatment Assumption (SITA)}: 

The treatment assignment $A$ is independent of potential outcomes $\left( Y^{(1)}, Y^{(0)} \right)$, given the covariates $\mathbf{X}$. 

\item[(C3)] $0 < \pi \left(\boldsymbol{x} \right) < 1$.

\item[(C4)] $(A,\mathbf{X})$ is independent and identically distributed.

\item[(C5)] $\mathbf{X}$ is bounded.

\end{itemize}
Conditions (C1)-(C5) allow us to characterize the distribution function $F^{(a)}$ by the variables $Y, A$ and $\bf{X}$ for $a\in\{0,1\}$. Specifically, let $I(\cdot)$ denote the indicator function. Then for $a=1$, we have that
\begin{equation}\label{identify}
   \begin{aligned}
        E\Bigg\{\frac{AI(Y\leq y)}{\pi(\bold{X})}\Bigg\} =&  E\Bigg[\frac{1}{\pi(\bold{X})}E\bigg\{AI(Y\leq y)\mid \bold{X}\bigg\}\Bigg] \\
        =&E\Bigg[\dfrac{E\bigg\{1\times I(Y\leq y)\mid \bold{X},A=1\bigg\}P(A=1\mid\bold{X})}{\pi(\bold{X})}\Bigg]\\
        &+ 
        E\Bigg[\dfrac{E\bigg\{0\times I(Y\leq y)\mid \bold{X},A=0\bigg\}P(A=0\mid\bold{X})}{\pi(\bold{X})}\Bigg] \\
        =&E\Bigg[\dfrac{E\bigg\{I(AY^{(1)}+(1-A)Y^{(0)}\leq y)\mid \bold{X},A=1\bigg\}\pi(\bold{X})}{\pi(\bold{X})}\Bigg]\\
        =&E\Bigg[{E\bigg\{I(Y^{(1)}\leq y)\mid \bold{X},A=1\bigg\}}\Bigg]\\
        =&E\Bigg[{E\bigg\{I(Y^{(1)}\leq y)\mid \bold{X}\bigg\}}\Bigg]\\
        =&E\bigg\{I(Y^{(1)}\leq y)\bigg\}\\
       =&F^{(1)}(y),
   \end{aligned}
\end{equation}
where the first equality is established from the law of iterated expectations, the second equality comes from {partitioning the conditional expectation $E\bigg\{AI(Y\leq y)\mid \bold{X}\bigg\}$ with the event $A=1$ and $A=0$, the third equality is a direct result of (C2) and $P(A=1\mid\bold{X}) = \pi(\bold{X})$, the fourth equality is obtained with basic operation, the fifth step utilizes (C2) again, and the sixth equality is derived from the law of iterated expectations with the inverse direction.} The identification procedure of $F^{(0)}(y)$ can be obtained by (\ref{identify}) with $A$ and $\pi(\bold{X})$  replaced by $1-A$ and $1-\pi(\bold{X})$, respectively. Consequently, under the independent and identically distributed (IID) samples with size $n$, denoted as  $\mathcal{O} \triangleq \bigl\{\{Y_i,A_i,\bold{X}_i\}: i=1,\dots,n\bigl\}$, with $\pi(\bold{X})$ being estimated consistently, the estimators of $F^{(1)}(y)$ and $F^{(0)}(y)$ are respectively given by
\begin{equation}\label{est-general treatment effect}
    \widehat{F}^{(1)}(y) = \bigg(\sum_{i=1}^n\dfrac{A_i}{\widehat{\pi}(\bold{X}_i)}\bigg)^{-1} \sum_{i=1}^n\dfrac{A_i I(Y_i\leq y)}{\widehat{\pi}(\bold{X}_i)} 
\end{equation}
and
\begin{equation}\label{9}
    \widehat{F}^{(0)}(y) =\bigg(\sum_{i=1}^n\dfrac{1-A_i}{1-\widehat{\pi}(\bold{X}_i)}\bigg)^{-1} \sum_{i=1}^n\dfrac{(1-A_i) I(Y_i\leq y)}{\widehat{\pi}(1-\bold{X}_i)}.
\end{equation}
The first notable remark for (\ref{est-general treatment effect}) and (\ref{9}) is that we implement $\bigg(\sum\limits_{i=1}^n\dfrac{A_i}{\widehat{\pi}(\bold{X}_i)}\bigg)$ and $\bigg(\sum\limits_{i=1}^n\dfrac{1-A_i}{1-\widehat{\pi}(\bold{X}_i)}\bigg)$ to replace $n$ when estimating the expectation (\ref{identify}) empirically because (\ref{est-general treatment effect}) and (\ref{9}) form the estimated  CDF; detailed justification can be found in Appendix \hyperref[justification]{C} of the supporting information. While Donald and Hsu (2014) also adopted the similar rationales in (\ref{est-general treatment effect}) and (\ref{9}), the difference is that our approach is for the estimation of CDF. 
\par
The other notable remark is that we focus on a scenario that $Y^{(a)}$ is continuous for $a\in\{0,1\}$ in the current study since our development is based on CDF. When $Y^{(a)}$ is discrete, the estimation problem naturally reduces to finding the location of largest jump in counterfactual CDFs. 
% using LD form -> mathematical reasons.
\subsection{Network Structure in Covariates}\label{Graph}

According to the structure in Figure~\ref{Causal-Network-fig}, the covariates $\bold{X}$ are possibly formulated by the network structure. To characterize the dependence structure for $\mathbf{X}$, we use an undirected \textit{graph}, denoted $\widetilde{G}\triangleq(\widetilde{V},\widetilde{E})$, where $\widetilde{V} = \{1,\cdots,p\}$ records the indexes of all components of $\mathbf{X}$, and $\widetilde{E} \subset \widetilde{V} \times \widetilde{V}$ contains dependent pairs of covariates. Following Chen and Yi (2021), we consider the exponential family graphical model that is formulated as follows:

\begin{eqnarray} \label{Exp_fam}
P(\mathbf{X};\boldsymbol{\beta},\boldsymbol{\Theta}) = \exp \left\{ \sum \limits_{r \in \widetilde{V}} \beta_r B(X_r)  + \sum \limits_{(s,\nu) \in \widetilde{E}} \theta_{s\nu} B(X_s)B(X_\nu) +  \sum \limits_{r \in \widetilde{V}} C(X_r) - \varphi(\boldsymbol{\beta},\boldsymbol{\Theta})  \right\}, 
\end{eqnarray}
where $\boldsymbol{\beta} = \left( \beta_1 , \cdots, \beta_p \right)^\top$ is a $p$-dimensional parameter vector, $\boldsymbol{\Theta} = \left[\theta_{s\nu}\right]$ is a $p \times p$ symmetric matrix, and $B(\boldsymbol{\cdot})$ and $C(\boldsymbol{\cdot})$ are given functions. The function $\varphi(\boldsymbol{\beta},\boldsymbol{\Theta})$ is the normalizing constant that makes (\ref{Exp_fam}) be integrated as 1.  For $r \in \widetilde{V}$, the parameter $\beta_r$ reflects the main effect associated with the covariate $X_r$; for $(s,\nu) \in \widetilde{E}$, the parameter $\theta_{s\nu}$ facilitates the association of $X_s$ and $X_\nu$ in the sense that $\theta_{s\nu} \neq 0$ shows the {\em conditional dependence} of $X_s$ and $X_\nu$ given other covariates.

\section{Methodology}\label{Sec3}

\subsection{Variable Selection and Network Construction for Outcome Responses}\label{variable selection}
Since the dataset contains multivariate confounders and some of them are not necessarily informative to estimate $\pi(\cdot)$ in (\ref{est-general treatment effect}) and (\ref{9}) and the treatment effects in (\ref{general treatment effect}), it is crucial to determine the important variables and detect the network structures that are dependent on the outcome response $Y$ (e.g., Shortreed and Ertefaie 2017). Let $V = \left\{r: \beta_r \neq 0, r=1,\cdots,p \right\}$ denote the indices set containing informative covariates, and let $E = \left\{(s,\nu): \theta_{s\nu} \neq 0 \right\}$ be the set containing the pairwise dependence, so that $G\triangleq(V,E)$. For the convenience of the presentation, we define $\mathbb{X}$ as the $n\times p$ matrix of confounders with $\mathbb{X}_j$ being the $j$th column vector in $\mathbb{X}$ for $j=1,\dots,p$, and let $\mathbb{X}^{\circ 2} \triangleq [\mathbb{X}_1* \mathbb{X}_2\; \mathbb{X}_1*\mathbb{X}_3\dots \mathbb{X}_{p-1}*\mathbb{X}_p]$, where $*$ is the entry-wise product of two column vectors. In addition, define two vectors $\bold{Y} = (Y_1\;Y_2\dots Y_n)^\top$ and $\boldsymbol{\varepsilon} = (\varepsilon_1\;\varepsilon_2\dots \varepsilon_n)^\top$ with $\varepsilon_i\sim\text{N}(0,\sigma^2)$, and denote $\text{vec}(\boldsymbol{\Theta})$ as the column vectorization of $\boldsymbol{\Theta}$. Then the matrix form  can be expressed as 
\begin{eqnarray}
    \bold{Y} = \mathbb{X}\boldsymbol{\beta}+ \mathbb{X}^{\circ2}\text{vec}(\boldsymbol{\Theta}) + \boldsymbol{\varepsilon}.
\end{eqnarray}
To select informative confounders and detect the corresponding network structure, we adopt the penalized likelihood method, and the estimators of $\boldsymbol{\beta}$ and $\boldsymbol{\Theta}$ are determined by
\begin{eqnarray}\label{object-min}
(\widehat{\boldsymbol{\beta}},\widehat{\boldsymbol{\Theta}}) &= \underset{\boldsymbol{\beta},\boldsymbol{\Theta}}{\argmin}\; \Bigg\{|| \bold{Y} - \mathbb{X}\boldsymbol{\beta} - \mathbb{X}^{\circ 2}\text{vec}(\boldsymbol{\Theta})||_2^2 + \lambda_1 || \boldsymbol{\beta}||_1 + \lambda_2 || \text{vec}(\boldsymbol{\Theta})||_1  \Bigg\},
\end{eqnarray}
where $||\cdot||_2$ is the $L_2$-norm and $\lambda_1$ and $\lambda_2$ are the tuning parameters associated with of the penalty terms $||\cdot||_1$ in the $L_1$-norm. Moreover, when $\widehat{\boldsymbol{\beta}}$ and $\widehat{\boldsymbol{\Theta}}$ are obtained, we define $\widehat{V} = \left\{r: \widehat{\beta}_r \neq 0 \right\}$ and $\widehat{E} = \left\{(s,\nu): \widehat{\theta}_{s\nu} \neq 0 \right\}$ as the corresponding estimated sets of $V$ and $E$, respectively, which reflect informative main effects and pairwise interaction with respect to $\bold{Y}$. Accordingly, the estimated undirected graph is defined as $\widehat{G} = (\widehat{V},\widehat{E})$. Since $ || \bold{Y} - \mathbb{X}\boldsymbol{\beta} - \mathbb{X}^{\circ 2}\text{vec}(\boldsymbol{\Theta})||_2^2$ and two penalty terms $|| \boldsymbol{\beta}||_1$ and $|| \text{vec}(\boldsymbol{\Theta})||_1$ are convex, then the objective function in (\ref{object-min}) is convex as well, ensuring the existence of the minimizer. To compute (\ref{object-min}), we adopt the coordinate-descent method to iteratively calculate the numerical values (e.g., Chen and Yi 2021).

\subsection{Estimation of Propensity Scores}\label{Sec 3.2}
In this section, we estimate the propensity score with selected confounders and the network structure taken into account.

To model the propensity score $\pi(\boldsymbol{x})$ in (\ref{PS}), we consider the following {\it network-based logistic regression model}:
\begin{equation}\label{logistic-regression}
    \text{logit}\left\{\pi(\boldsymbol{x})\right\} = \eta_0 + \sum \limits_{r \in \widehat{V}} \eta_r x_r + \sum \limits_{(s,\nu) \in \widehat{E}} \eta_{s\nu} x_s x_\nu, 
\end{equation}
where $\eta_0$ is an intercept, $\sum \limits_{r \in \widehat{V}} \eta_r x_r$ reflects the informative confounders that are highly correlated with outcomes, and $\sum \limits_{(s,\nu) \in \widehat{E}} \eta_{s\nu} x_s x_\nu$ reflects the dependence structures among confounders. Let $\boldsymbol{\eta} \triangleq \left( \eta_0, \boldsymbol{\eta}_V^\top, \boldsymbol{\eta}_E^\top \right)^\top$, where $\boldsymbol{\eta}_V \triangleq \left( \eta_r : r \in \widehat{V} \right)^\top$ and $\boldsymbol{\eta}_E \triangleq \left( \eta_{s\nu} : (s,\nu) \in \widehat{E} \right)^\top$ are two vectors of parameters. Based on the sample $\mathcal{O}$, we apply the maximum likelihood method based on logistic regression model to estimate  $\boldsymbol{\eta}$, and the estimator is given by
\begin{equation}\label{logistic-estimator}
    \widehat{\boldsymbol{\eta}} = \argmax \limits_{\boldsymbol{\eta}} \left[ \prod \limits_{i=1}^n \left\{ \pi(\boldsymbol{x}_i) \right\}^{a_i} \left\{ 1 - \pi(\boldsymbol{x}_i) \right\}^{1-a_i} \right].
\end{equation}
Therefore, the proposed estimator of the propensity score is given by
\begin{eqnarray} \label{est-PS}
\widehat{\pi}(\boldsymbol{x}) = \frac{\exp\left( \widehat{\eta}_0 + \sum \limits_{j \in \widehat{V}} \widehat{\eta}_r x_r + \sum \limits_{(s,\nu) \in \widehat{E}} \widehat{\eta}_{s\nu} x_s x_\nu \right)}{1 + \exp\left( \widehat{\eta}_0 + \sum \limits_{r \in \widehat{V}} \widehat{\eta}_r x_r + \sum \limits_{(s,\nu) \in \widehat{E}} \widehat{\eta}_{s\nu} x_s x_\nu \right)}.
\end{eqnarray}

(\ref{est-PS}) is an extended formulation from the conventional logistic regression models by incorporating network structures. This expression is also similar to Chen et al. (2019), which is numerically justified that incorporating network structures would improve the accuracy of the classification.

\subsection{Estimation of IPW-CDFs}
When propensity scores $\pi(\cdot)$ are estimated by (\ref{est-PS}), $F^{(a)}$ is estimated by (\ref{est-general treatment effect}) and (\ref{9}) for $a\in\{0,1\}$. Consequently, the estimator of (\ref{general treatment effect}) is given by 
\begin{equation}\label{est-general treatment effect2}
    T(\widehat{F}^{(1)})-T(\widehat{F}^{(0)}),
\end{equation}
which reduces to the estimators of various treatment effects in Section \ref{Sec2.1}. For example, the ATE (\ref{ATE}) is estimated by
\begin{eqnarray}\label{est-ATE}
    \widehat{\tau} = \int_{\mathbb{R}} y d\widehat{F}^{(1)}(y) - \int_{\mathbb{R}} y d\widehat{F}^{(0)}(y).
\end{eqnarray}
The estimators of the $q$th quantiles in (\ref{quantile1}) are given by
\begin{eqnarray}\label{est-quantile1}
    \widehat{\xi}^{(1)}(q) = \inf\bigg\{y\in \mathbb{R}\mid \widehat{F}^{(1)}(y)\geq q\bigg\} \text{ and }\widehat{\xi}^{(0)}(q) = \inf\bigg\{y\in \mathbb{R}\mid \widehat{F}^{(0)}(y)\geq q\bigg\},
\end{eqnarray}
yielding the estimator of~(\ref{QTE})
\begin{eqnarray}\label{est-QTE}
    \widehat{\Xi}(q) =  \widehat{\xi}^{(1)}(q) -  \widehat{\xi}^{(0)}(q)
\end{eqnarray}
for some $q\in(0,1)$. Finally, the estimator of DTE (\ref{DTE}) is given by the difference of (\ref{est-general treatment effect}) and (\ref{9}), i.e,
\begin{eqnarray}\label{est-DTE}
    \widehat{\Delta}(y) = \widehat{F}^{(1)}(y) - \widehat{F}^{(0)}(y)
\end{eqnarray}
for a given $y\in\mathbb{R}$.
\par
Here we comment the proposed estimator (\ref{est-general treatment effect2}). The first advantage of (\ref{est-general treatment effect2}) is robustness. Specifically, as pointed out by Khan and Ugander (2023), $\bigg(\sum\limits_{i=1}^n\dfrac{A_i}{\widehat{\pi}(\boldsymbol{X}_i)}\bigg)$ and $\bigg(\sum\limits_{i=1}^n\dfrac{1-A_i}{1-\widehat{\pi}(\boldsymbol{X}_i)}\bigg)$ in the denominator of the estimators  (\ref{est-general treatment effect}) and (\ref{9}) are positively correlated with the numerators, the effect of the extreme values can be mitigated by such correlation. In addition, (\ref{est-general treatment effect2}) is formulated by the CDF that lie in an interval $[0,1]$, which can also avoid potential impact of outliers in the outcome $Y$. The second advantage of the proposed estimator (\ref{est-general treatment effect2}) is to provide a general formulation that unifies a class of various causal effects and incorporates variable selection to avoid unnecessary bias indcued by redundant covariates and pairwise dependence structure. When variable selection is not taken into account, (\ref{est-ATE}) reduces to the estimator described in Lunceford and Davidan (2004). Furthermore, if the treatment model is assigned as series logit model (Hirano, Imbens and Ridder, 2003) with variable selection ignored, (\ref{est-QTE}) reduces to the estimator proposed by Donald and Hsu (2014). Finally, the estimator (\ref{est-DTE}) can be referred to the generalization of caused effects of survivor functions.

\subsection{Theoretical Results }\label{Theory}
To establish the asymptotic properties, we first present the consistency of variable selection in the following theorem, which ensures that the truly informative confounders and the network structure can be identified.
\begin{theorem}\label{recovery}
    Under regularity conditions stated in the Supporting Information, the following property holds:
    $$
    P(\widehat{G}= G) \to 1\text{ as }n\to\infty.
    $$ 
\end{theorem}
Theorem \ref{recovery} states that there is a high probability that our estimation procedure can successfully capture the underlying true confounders and dependence structure among covariates for the potential outcome. Consequently, our estimators are free of the potential bias incurred by non-informative confounders.

Before introducing the asymptotic properties for the proposed estimators, we first define some notations. We define the density functions of $Y^{(1)}$ and $Y^{(0)}$ as $f^{(1)}$ and $f^{(0)}$, respectively. With Theorem \ref{recovery}, we are capable of using the selected covariates. Let $\bold{X}_{\widehat{V}}$ and $\bold{X}_{\widehat{E}}$ be two matrices with dimensions $n\times k$ and $n\times m$ consisting of the chosen main and interaction effects, respectively. Consequently, we define $\bold{X}^* =\bigg[\bold{1}\;\bold{X}_{\widehat{V}}\;\bold{X}_{\widehat{E}}\bigg]$ where $\bold{1}$ is the column vector with all components being $1$.

To derive the theoretical properties, we further impose some assumptions to the statistical functional. Suppose that $T:\mathcal{F}\to\mathbb{R}$ satisfies Hadamard differentiability at $F\in\mathcal{F}$, where $\mathcal{F}$ is a normed space consisting of probability distribution functions, i.e., there exists a continuous linear function $T^\prime_F$, such that, as $n\to\infty$, 
$$
\dfrac{T(F+t_n h_n)-T(F)}{t_n}\to T^\prime_F(h)
$$
{for all sequences $\{t_n\}\subset\mathbb{R}$ and $\{h_n\}\subset\mathcal{F}$ satisfying $t_n\to0$, $h_n\to h\in\mathcal{F}$ and $F+t_nh_n\in\mathcal{F}$.} In addition, when the statistical functional $T$ is Hadamard differentiable, the Gateaux derivative, which is also called the von Mises derivative, of $T$ exists and the values of these two types of derivative are the same (van der Vaart, 2000). Furthermore, the influence curve, or the influence function, of a statistical functional $T$ is defined as
\begin{equation}\label{influence curve}
    \phi_F(y) \triangleq \lim_{t\to0}\frac{T((1-t)F + t\delta_y) - T(F)}{t},
\end{equation}
where $\delta_y$ is the distribution function with point mass $1$ on $y$ (Fernholz, 1983). In particular, if a statistical functional can be written as the form $\int \varphi(y)dF(y)$ for some function $\varphi(y)$, it is called a linear functional and its influence curve is $\phi_F(x) = \varphi(x) - T(F)$ (Wasserman, 2006).

We now present the asymptotic properties of the proposed estimator.
\begin{theorem}\label{asymptotic for our estimator}
Under regularity conditions stated in the Supporting Information, the following properties hold as $n\to\infty$:
\begin{itemize}
    \item[(i)]$
\sqrt{n}(\widehat{\boldsymbol{\eta}} - \boldsymbol{\eta})\overset{d}{\longrightarrow}\text{N}(\bold{0},\mathcal{A}^{-1});
    $
    \item[(ii)] $      \sqrt{n}\Bigg[\bigg(T(\widehat{F}^{(1)})-T(\widehat{F}^{(0)})\bigg)-\bigg(T({F}^{(1)})-T({F}^{(0)})\bigg)\Bigg]\overset{d}{\longrightarrow}\text{N}\bigg(0, -\mathcal{B}\mathcal{A}^{-1}\mathcal{B}^\top +\mathcal{C}\bigg)$,
\end{itemize}
where
$$
\mathcal{A}= E\Bigg\{\pi(\bold{X}^*\mid\boldsymbol{\eta})\bigg(1-\pi(\bold{X}^*\mid\boldsymbol{\eta})\bigg) \bold{X}^{*\top}\bold{X}^{*}\Bigg\},
$$
$$
\begin{aligned}
    \mathcal{B} =E\Bigg[\bigg\{\dfrac{\phi_{F^{(1)}}(Y^{(1)})}{\pi(\bold{X}^*\mid\boldsymbol{\eta})}+\dfrac{\phi_{F^{(0)}}(Y^{(0)})}{1-\pi(\bold{X}^*\mid\boldsymbol{\eta})}\bigg\}\frac{\partial}{\partial\boldsymbol{\eta}} \pi(\bold{X}^*\mid\boldsymbol{\eta})\Bigg]
\end{aligned}
$$
and
  $$
  \mathcal{C} =E\Bigg\{\dfrac{\phi^2_{F^{(1)}}(Y^{(1)})}{\pi(\bold{X}^*\mid\boldsymbol{\eta})}+\dfrac{\phi^2_{F^{(0)}}(Y^{(0)})}{1-\pi(\bold{X}^*\mid\boldsymbol{\eta})}\Bigg\}.
  $$ 
\end{theorem}
Theorem \ref{asymptotic for our estimator} (i) shows the asymptotic normality of the logistic estimates derived in Section \ref{Sec 3.2}, which ensures the validity of the treatment model throughout the estimation procedure. Theorem \ref{asymptotic for our estimator} (ii) states that, when $n\to\infty$, the estimated treatment effect under a given statistical functional $T$ follows a normal distribution with sandwich type variance. With the propensity score being estimated, the estimated treatment effct $T(\widehat{F}^{(1)})-T(\widehat{F}^{(0)})$ is more efficient than that under the given propensity score, which is due to the additional term $-\mathcal{B}\mathcal{A}^{-1}\mathcal{B}^\top$ from the result (i). The novel contribution in Theorem \ref{asymptotic for our estimator} (ii) is that the treatment effect holds for any statistical functional $T(\cdot)$. To see the generality of Theorem \ref{asymptotic for our estimator}, we revisit (\ref{est-ATE}), (\ref{est-QTE}) and (\ref{est-DTE}), and connect these theoretical results with Theorem \ref{asymptotic for our estimator} in the following discussion.
\begin{Exa}[Average Treatment Effect]\label{example-ate}
 Let $T(F) = \int x dF(x)$ with $\varphi(y)=y$, then (\ref{est-general treatment effect2}) reduces to (\ref{est-ATE}).  Thus, the influence curve is $\phi_F(y) = y-T(F)$. By Theorem \ref{asymptotic for our estimator} (ii), we have that, as $n\to\infty$,
 $$
 \sqrt{n}(\widehat{\tau}-\tau) \overset{d}{\longrightarrow} \text{N}\bigg(0, -\mathcal{B}_{\textup{ATE}}\mathcal{A}^{-1}\mathcal{B}_{\textup{ATE}}^\top +\mathcal{C}_{\textup{ATE}} \bigg),
 $$
 where the components of the asymptotic variance are given by
$$
\mathcal{B}_{\textup{ATE}}= E\Bigg[\bigg\{\dfrac{Y^{(1)}-E(Y^{(1)})}{\pi(\bold{X}^*\mid\boldsymbol{\eta})}+\dfrac{Y^{(0)}-E(Y^{(0)})}{1-\pi(\bold{X}^*\mid\boldsymbol{\eta})}\bigg\}\frac{\partial}{\partial\boldsymbol{\eta}} \pi(\bold{X}^*\mid\boldsymbol{\eta})\Bigg]
$$
 and
 $$
\mathcal{C}_{\textup{ATE}} =E\bigg[\dfrac{\{Y^{(1)}-E(Y^{(1)})\}^2}{\pi(\bold{X}^*\mid\boldsymbol{\eta})}+\dfrac{\{Y^{(0)}-E(Y^{(0)})\}^2}{1-\pi(\bold{X}^*\mid\boldsymbol{\eta})}\bigg].
 $$
This the same as that in Lunceford and Davidian (2004).
\end{Exa}

\begin{Exa}[Quantile Treatment Effect]
   Let $T(F) = \inf\{y: F(y) \ge q \}\triangleq \xi(q)$ for some $0<q<1$, then (\ref{est-general treatment effect2}) reduces to the estimator (\ref{est-QTE}). By Wasserman (2006, p.21), the influence curve is given by $\phi_F(y) = \dfrac{q-I_{(-\infty,\xi(q)]}(y)}{f(\xi(q))}$, where $f$ is the density function of $F$. Theorem \ref{asymptotic for our estimator} (ii) is reduced to
   $$
   \sqrt{n}\bigg(\widehat{\Xi}(q)-\Xi(q)\bigg) \overset{d}{\longrightarrow} \text{N}\bigg(0, -\mathcal{B}_{\textup{QTE}}\mathcal{A}^{-1}\mathcal{B}_{\textup{QTE}}^\top +\mathcal{C}_{\textup{QTE}}  \bigg)
   $$
   as $n\to\infty$, where the components of the asymptotic variance are given by
  $$
\mathcal{B}_{\textup{QTE}} = E\Bigg[ \bigg\{\dfrac{q-I_{(-\infty,\xi^{(1)}(q)]}(Y^{(1)})}{f^{(1)}(\xi^{(1)}(q))\pi(\bold{X}^*\mid\boldsymbol{\eta})}+\dfrac{q-I_{(-\infty,\xi^{(0)}(q)]}(Y^{(0)})}{f^{(0)}(\xi^{(0)}(q))[1-\pi(\bold{X}^*\mid\boldsymbol{\eta})]}\bigg\}\frac{\partial}{\partial\boldsymbol{\eta}} \pi(\bold{X}^*\mid\boldsymbol{\eta})\Bigg]
$$
 and
 $$
 \mathcal{C}_{\textup{QTE}} =\frac{1}{f^{(1)}(\xi^{(1)}(q))^2}E\bigg[\dfrac{(q-I_{(-\infty,\xi^{(1)}(q)]}(Y^{(1)}))^2}{\pi(\bold{X}^*\mid\boldsymbol{\eta})}\bigg]+\frac{1}{f^{(0)}(\xi^{(0)}(q))^2}E\bigg[\dfrac{(q-I_{(-\infty,\xi^{(0)}(q)]}(Y^{(0)}))^2}{1-\pi(\bold{X}^*\mid\boldsymbol{\eta})}\bigg].
 $$
 Note that $\frac{1}{f^{(1)}(\xi^{(1)}(q))^2}E\bigg[\dfrac{(q-I_{(-\infty,\xi^{(1)}(q)]}(Y^{(1)}))^2}{\pi(\bold{X}^*\mid\boldsymbol{\eta})}\bigg]$ and $\frac{1}{f^{(0)}(\xi^{(0)}(q))^2}E\bigg[\dfrac{(q-I_{(-\infty,\xi^{(0)}(q)]}(Y^{(0)}))^2}{1-\pi(\bold{X}^*\mid\boldsymbol{\eta})}\bigg]$ coincide with the similar form of the semiparametric efficiency bound derived by Firpo (2007). The difference in the forms of asymptotic variance results from the different approaches to the estimation of propensity score.  To model the relationship between treatment and confounders, we employ logistic regression whereas Firpo (2007) adopt the logit series approximation from Hirano, Imbens and Ridder (2003).
\end{Exa}
\begin{Exa}[Distributional Treatment Effect] %%% For asymptotic normality, see Proposition 6.1.8 in Fernholz (1983)
Let $T(F) = \int I_{(-\infty,y]}(x) dF(x) = F(y)$ with $\varphi(x) = I_{(-\infty,y]}(x)$, then (\ref{est-general treatment effect2}) reduces to (\ref{est-DTE}).  The resulting influence curve is $\phi_F(x) = I_{(-\infty,y]}(x) - T(F)$. According to Theorem \ref{asymptotic for our estimator} (ii), we have that, as $n\to\infty$,
 $$
 \sqrt{n}\bigg(\widehat{\Delta}(y)-\Delta(y)\bigg) \overset{d}{\longrightarrow} \text{N}\bigg(0,-\mathcal{B}_{\textup{DTE}}\mathcal{A}^{-1}\mathcal{B}_{\textup{DTE}}^\top +\mathcal{C}_{\textup{DTE}}   \bigg),
 $$
 where the components of the asymptotic variance are given by
$$
\mathcal{B}_{\textup{DTE}} = E\Bigg[\bigg\{\dfrac{I_{(-\infty,y]}(Y^{(1)})-F^{(1)}(y)}{\pi(\bold{X}^*\mid\boldsymbol{\eta})}+\dfrac{I_{(-\infty,y]}(Y^{(0)})-F^{(0)}(y)}{1-\pi(\bold{X}^*\mid\boldsymbol{\eta})}\bigg\}\frac{\partial}{\partial\boldsymbol{\eta}} \pi(\bold{X}^*\mid\boldsymbol{\eta})\Bigg]
$$
 and
 $$
 \mathcal{C}_{\textup{DTE}}  =E\bigg[\dfrac{\{I_{(-\infty,y]}(Y^{(1)})-F^{(1)}(y)\}^2}{\pi(\bold{X}^*\mid\boldsymbol{\eta})}+\dfrac{\{I_{(-\infty,y]}(Y^{(0)})-F^{(0)}(y)\}^2}{1-\pi(\bold{X}^*\mid\boldsymbol{\eta})}\bigg].
 $$
\end{Exa}

\section{Numerical Studies}\label{Sec4}
In this section, we design a series of synthetic data to assess the performance of the proposed method. To show the advantage of the proposed method, we also examine several existing methods.
\subsection{Simulation Design}
Let the dimension of covariates be $p=12$ and let $\boldsymbol{x}$ denote the $p$-dimensional confounders. We consider four different sample sizes $n=500, 1000,2000$ or $10000$. In the numerical studies, we design two different scenarios to generate the covariates. In Scenario 1, we consider the independent covariates and independently generate $\boldsymbol{x}$ by the multivariate normal distribution with mean zero and identity covariance matrix, which implies that there is no network structure among confounders.

In Scenario 2, we consider the dependent covariates, which are generated by the multivariate normal distribution with mean zero and the covariance matrix reflecting two network structures in Figure~\ref{Sim-Network-fig}.

When $\mathbf{X}$ is generated, we respectively generate $A$ and $Y$ by the following two models
\begin{equation}\label{model A}
\text{logit}\left\{ P(A=1|\mathbf{X}) \right\} = 1 + X_1 + X_3 + \sum_{(s,\nu)\in {E}} X_SX_{\nu}
\end{equation}
and
\begin{equation}\label{model Y}
Y = \gamma_0 A + 1 + X_1 + X_3 + \sum_{(s,\nu)\in{E} }X_SX_{\nu}+\varepsilon,
\end{equation}
where $\varepsilon \sim N(0,1)$, $\gamma_0 = 1$ and $E$ is the set that is either empty under Scenario 1 or contains pairs in Scenario 2. From (\ref{model A}) and (\ref{model Y}), variables $X_1$ and $X_3$ and pairs in ${E}$ are confounders. Consequently, we obtain the IID sample $\mathcal{O} = \bigg\{ \{ Y_i,A_i,\bold{X}_i:i=1,\dots,n\}$ with size $n$.

In this study, we primarily examine the estimators (\ref{est-ATE}), (\ref{est-QTE}) with $q\in\{0.2,0.25,0.5,0.75,0.8\}$, and  (\ref{est-DTE}) with $y\in\{-3,0,3\}$. We compare the proposed method with existing approaches. For the estimation of the ATE, we primarily examine the inverse probability weighted estimator under the outcome $Y_i$
\begin{eqnarray}\label{ATE-IPW1}
    \tau_{\text{IPW}} = \frac{1}{n}\bigg(\sum_{i=1}^n\frac{A_iY_i}{\widehat{\pi}(\bold{X}_i)}\bigg) - \frac{1}{n}\bigg(\sum_{i=1}^n\frac{(1-A_i)Y_i}{1-\widehat{\pi}(\bold{X}_i)}\bigg)
\end{eqnarray}
and the modified version proposed by Lunceford and Davidian (2004)
\begin{eqnarray}\label{ATE-IPW2}
        \tau_{\text{LD}} = \bigg(\sum_{i=1}^n\frac{A_i}{\widehat{\pi}(\bold{X}_i)}\bigg)^{-1}\bigg(\sum_{i=1}^n\frac{A_iY_i}{\widehat{\pi}(\bold{X}_i)}\bigg) - \bigg(\sum_{i=1}^n\frac{1-A_i}{1-\widehat{\pi}(\bold{X}_i)}\bigg)^{-1}\bigg(\sum_{i=1}^n\frac{(1-A_i)Y_i}{1-\widehat{\pi}(\bold{X}_i)}\bigg),
\end{eqnarray}
where (\ref{ATE-IPW1}) and (\ref{ATE-IPW2}) do not take variable selection into account.

For the estimation of the QTE, we examine the approach proposed in Firpo (2007), which is given by
\begin{eqnarray}\label{QTE-Firpo}
    \widehat{\Xi}_{\text{Firpo}}(q) = \widehat{\xi}_{\text{Firpo}}^{(1)}(q)-\widehat{\xi}_{\text{Firpo}}^{(0)}(q),
\end{eqnarray}
where
$$
    \widehat{\xi}_{\text{Firpo}}^{(0)}(q) = \underset{q}{\text{argmin}} \frac{1}{n} \sum_{i=1}^{n} \frac{1-A_i}{1-\widehat{\pi}(\bold{X}_i)}\cdot\rho_q(Y_i-q)\text{ and }\widehat{\xi}_{\text{Firpo}}^{(1)}(q)= \underset{q}{\text{argmin}} \frac{1}{n} \sum_{i=1}^{n} \frac{A_i}{\widehat{\pi}(\bold{X}_i)}\cdot\rho_q(Y_i-q)
$$
with $\rho_q$ being the check function stated in Koenker (2005). Noting that the estimators (\ref{est-QTE}) and (\ref{QTE-Firpo}) form a group of comparison, where the former incorporates variable selection in the estimation procedure while the latter does not.

To assess the performance of variable selection, we compute the specificity (SPE) and sensitivity (SEN), which are respectively given by
$$
\text{SEN} =\dfrac{\#\bigg\{j,(s,\nu):\theta_{s\nu} = 0, \widehat{\theta}_{s\nu}=0,\beta_j =0,\widehat{\beta}_j=0\bigg\}}{\#\bigg\{j,(s,\nu):\theta_{s\nu} = 0,\beta_j=0\bigg\}} 
$$
and
$$
\text{SPE}=\dfrac{\#\bigg\{j,(s,\nu):\theta_{s\nu} \not= 0, \widehat{\theta}_{s\nu}\not=0,\beta_j\not=0,\widehat{\beta}_j\not=0\bigg\}}{\#\bigg\{j,(s,\nu):\theta_{s\nu} \not= 0, \beta_j\not=0\bigg\}}.
$$
In addition, to examine the performance of the estimators derived by the proposed or existing methods, we compute the biases (BIAS), standard errors (S.E.), mean squared errors (MSE) and coverage rate (CR). For each setting,  we repeat the simulation {1000} times.
\subsection{Simulation Results}
The simulation results of ATE, QTE and DTE under Scenario 1 are summarized in Tables \ref{ATE-simulation-sc1}, \ref{QTE-simulation-sc1} and \ref{DTE-simulation-sc1}, respectively; the results under Scenario 2 are recorded in Tables \ref{ATE-simulation}, \ref{QTE-simulation} and \ref{DTE-simulation}, respectively. For variable selection, we find that both sensitivity and specificity are close or equal to one regardless of network structures, which indicate that the estimated graphical structure and selected informative confounders can be recovered to the true ones and are consistent with the finding in Theorem \ref{recovery}.

From the estimation of various treatments, our CDF approach generally outperforms the existing methods and is more efficient with smaller MSE. The CR of the proposed method is also close to 95\%. To see the performance of each treatment effects, we first observe from ATE that our method shows numerical equivalence to the method from Lunceford and Davidian (2004), which matches the finding in Example \ref{example-ate} in Section \ref{Theory}. In addition, the proposed and Lunceford and Davidian’s methods are more efficient than the traditional IPW method with smaller S.E. and MSE. When we look at the QTE, we find that our method outperforms Firpo's (2007) method with smaller bias and MSE even in Scenario 1. It indicates that the proposed CDF approach is robust. Finally, the performance for DTE shows that our method is capable of offering precise estimates due to the lower bias, S.E. and MSE and approximate 95\% CR.

\section{Analysis of NHEFS data}\label{CHILD}
In this section, we implement the proposed method in Section~\ref{Sec3} to analyze the data arising from NHANES I Epidemiologic Follow-up Study (NHEFS), which is a national longitudinal study collaboratively launched by the National Center for Health Statistics and the National Institute on Aging. The NHEFS aimed to analyze the links between clinical, nutritional, and behavioral variables. For the detailed description, one can refer to \url{https://wwwn.cdc.gov/nchs/nhanes/nhefs/default.aspx} for the official information.

The full data set, which is available at \url{https://miguelhernan.org/whatifbook}, contains 1430 subjects and several variables including weight measured in kilograms (\texttt{wt}), the status of smoking behavior (\texttt{qsmk}), systolic blood pressure (\texttt{sbp}, in millimeter of mercury (mmHg)), serum cholesterol (\texttt{cholesterol}, in mg/100), number of cigarettes smoked per day 
in 1971 (\texttt{smokeintensity}), diastolic blood pressure (\texttt{dbp}, in mmHg), height in centimeters (\texttt{ht}), average tobacco price in the state of residence in 1982 (\texttt{price82}, in US dollars), years of smoking (\texttt{smokeyrs}), age in 1971 (\texttt{age}, in years), and total family income in 1971 (\texttt{income}). We aim to explore the causal effects of smoking habits on the change of weight. In this study, we aim to explore various types of causal effects of weights from different status of smoking behavior. Following the notation from Section \ref{Sec2}, we take $Y$ and $A$ as \texttt{wt} and \texttt{qsmk}, respectively. Moreover, we let the remaining nine variables be confounders. Since those nine confounders may form pairwise dependence structure and part of them is informative to $Y$, it is crucial to implement variable selection technique in Section \ref{Sec3} and then estimate the treatment effects. In addition, to see the impact of variable selection, we apply the existing methods in Section \ref{Sec4} for the comparisons.

The proposed method suggests that the selected confounders are \texttt{ht} and \texttt{age}, and the network structure of nine confounders is displayed in Figure \ref{0.01}. The revealed dependence structure is centered around \texttt{race} and \texttt{sex} while \texttt{age} does not contribute to the dependence structure with other variables. This indicates that when it comes to smoking behavior and weight gain,  \texttt{race} and \texttt{sex} are two critical variables that possess further underlying connections with other covariates and should be taken into account while analyzing causality in the data.

Our interest lies in estimating (\ref{ATE}), (\ref{QTE}) with $q=0.2,0.25,0.5,0.75$ and $0.8$, and (\ref{DTE}) evaluated at $y=\Bar{Y}$, the sample average of the outcome. The estimation results of the proposed and existing methods are recorded in Table \ref{estimated causal effects}, where EST represents the estimator, S.E. is the standard error obtained by the bootstrap method with 10000 repetitions, and p-value is used to examine two-tailed hypotheses $H_0: \tau=0$, $H_0:\Xi(q)=0$ or $H_0:\Delta(y)=0$ for some pre-specified $q$ and $y$. We find all estimates are greater than 2 except for DTE. Consistent with the finding in simulation, the proposed CDF method is more efficient with smaller S.E. Under level of significance 0.05, all estimates are significant except for $\Xi(0.75)$, which indicate that smoking generally causes an increase in weight.  Regarding the results of $\Xi(0.75)$, we comment that Firpo's method might falsely reject the null hypothesis, i.e. $\text{H}_0:\;\Xi(0.75)=0$, as the simulation results suggested that Firpo's method rarely captures the true parameter from the perspective of CR.

\section{Summary} \label{Summary}
Treatment effect estimations have been an important issue in causal inference, which include but not are not limited to ATE, QTE and DTE. The other challenge for the estimation is the multivariate/high-dimensional data with potential network structure. To tackle these problems, we propose a flexible and unified methodology that is based on statistical functional structure and accommodates the estimation of various causal effects, including ATE, QTE, DTE. Our apporach is under the CDF approach and takes the detection of network structure among covariates into account. Theoretically, the variable selection consistency and asymptotic normality properties of the statisitcal functional estimator are rigorously established, which ensure the validity of the proposed estimator. Numerical studies show that the finite sample performance of the proposed method is satisfactory and even outperform other existing methods. 

The findings illustrate the framework’s flexibility in estimating causal effects. The proposed method offers a versatile and effective solution for exploring diverse causal measures in complex datasets. This unified approach represents a significant advance in the field of causal inference, providing researchers with a powerful tool for uncovering insights in high-dimensional data.

\section*{References}

\refmark Chapfuwa, P., Assaad, S., Zeng, S., Pencina, M. J., Carin, L. and Henao, R. (2021, April). Enabling counterfactual survival analysis with balanced representations. In \textit{Proceedings of the Conference on Health, Inference, and Learning} (pp. 133-145). %

\refmark Chen, L. P. (2020). Causal inference for left-truncated and right-censored data with covariate measurement error. \textit{Computational and Applied Mathematics}, 39, 1-27.%

\refmark Chen, L.-P. (2024). Estimation of graphical models: an overview of selected topics. {\em International Statistical Review}, 92: 194-245.

\refmark Chen, L.-P. and Yi, G. Y. (2021). Analysis of noisy survival data with graphical proportional hazards measurement error model. {\em Biometrics}, 77, 956-969. 

\refmark Chen, L.-P., Yi, G. Y., Zhang, Q. and He, W. (2019) Multiclass analysis and prediction with network  structured covariates. {\em Journal of Statistical Distributions and Applications}, 6:6.%

\refmark Chernozhukov, V., Fernández‐Val, I. and Melly, B. (2013). Inference on counterfactual distributions. \textit{Econometrica}, 81, 2205-2268.%

\refmark Donald, S. G. and Hsu, Y. C. (2014). Estimation and inference for distribution functions and quantile functions in treatment effect models. \textit{Journal of Econometrics}, 178, 383-397.%

\refmark Ertefaie, A., Asgharian, M. and Stephens, D. A. (2018). Variable selection in causal inference using a simultaneous penalization method. {\em Journal of Causal Inference}, 6, 20170010.%

\refmark Firpo, S. (2007). Efficient semiparametric estimation of quantile treatment effects. \textit{Econometrica}, 75, 259-276.%

\refmark{Hirano, K., Imbens, G. W. and Ridder, G. (2003). Efficient estimation of average treatment effects using the estimated propensity score. \textit{Econometrica}, 71, 1161-1189.}

\refmark Hsu, Y. C., Lai, T. C. and Lieli, R. P. (2022). Counterfactual treatment effects: Estimation and inference. \textit{Journal of Business \& Economic Statistics}, 40, 240-255.%

\refmark Khan, S. and Ugander, J. (2023). Adaptive normalization for IPW estimation. \textit{Journal of Causal Inference}, 11, 20220019.

\refmark Koch, B., Vock, D. M., Wolfson, J. and Vock, L. B. (2020). Variable selection and estimation in causal inference using Bayesian spike and slab priors. {\em Statistical Methods in Medical Research,} 29, 2445-2469. %

\refmark Koenker, R. (2005). {\em Quantile Regression}. Cambridge University Press.%

\refmark Lunceford, J. K. and Davidian, M. (2004). Stratification and weighting via the propensity score in estimation of causal treatment effects: a comparative study. {\em Statistics in Medicine}, 23, 2937-2960 %

\refmark Rosenbaum, P. R. and Rubin, D. B. (1983) The central role of the propensity score in observational studies for causal effects. {\em Biometrika}, {70}, 41-55. %

\refmark Shortreed, S. M. and Ertefaie, A. (2017). Outcome-adaptive lasso: variable selection for causal inference. {\em Biometrics}, 73, 1111-1122.%

\refmark{van der Vaart, A. W. (2000). {\em Asymptotic Statistics}. Cambridge university press.}

\refmark von Mises, R. (1947). On the asymptotic distribution of differentiable statistical functions. \textit{The Annals of Mathematical Statistics}, 18, 309-348.

\refmark Wasserman, L. (2006). {\em All of nonparametric statistics}. Springer Science \& Business Media.

\refmark Yi, G. Y. and Chen, L. P. (2023). Estimation of the average treatment effect with variable selection and measurement error simultaneously addressed for potential confounders. \textit{Statistical Methods in Medical Research}, 32, 691-711.%

\refmark Zou, H. (2006). The adaptive lasso and its oracle properties. {\em Journal of the American Statistical Association}, 101, 1418-1429.%

\clearpage
\begin{landscape}
% SPE: 1.000, CR: 0.000 do not cross page
    \begin{longtable}{clcccccc}
     \caption{Simulation results for ATE under Scenario 1}
     \renewcommand{\arraystretch}{0.7}
     \label{ATE-simulation-sc1}\\
\hline
$n$   & Method & SEN   & SPE & BIAS   & S.E.  & MSE   & CR    \\
$500$   
      & IPW    & -     & -   & 0.010  & 0.344 & 0.118 & 0.970 \\
      & LD     & -     & -   & 0.026  & 0.251 & 0.063 & 0.959 \\
      & CDF    & 0.998 & 1.000   & 0.026  & 0.251 & 0.063 & 0.959 \\ \hline

$1000$  
      & IPW    & -     & -   & 0.006  & 0.261 & 0.068 & 0.962 \\
      & LD     & -     & -   & 0.017  & 0.195 & 0.038 & 0.957 \\
      & CDF    & 0.999 & 1.000   & 0.017  & 0.195 & 0.038 & 0.957 \\ \hline

$2000$ 
      & IPW    & -     & -   & 0.002  & 0.175 & 0.030 & 0.963 \\
      & LD     & -     & -   & 0.007  & 0.138 & 0.019 & 0.961 \\
      & CDF    & 0.999 & 1.000   & 0.007  & 0.138 & 0.019 & 0.961 \\ \hline

$10000$ 
      & IPW    & -     & -   & -0.000 & 0.088 & 0.007 & 0.969 \\
      & LD     & -     & -   & 0.000  & 0.070 & 0.005 & 0.968 \\
      & CDF    & 1.000     & 1.000   & 0.000  & 0.070 & 0.005 & 0.968 \\ \hline
\end{longtable}
\end{landscape}

\begin{landscape}
    \renewcommand{\arraystretch}{0.92}
    \begin{longtable}{cllcccccc}
    \caption{Simulation results for QTE under Scenario 1}
   
    \label{QTE-simulation-sc1}\\
     \hline
$n$                  & Estimand              & Method & SEN   & SPE & BIAS   & S.E.  & MSE   & CR    \\
$500$                & $\Xi(0.2)$           
                      & Firpo  & -     & -   & 1.457  & 0.209 & 2.124 & 0.000     \\
                     &  & CDF    & 0.998 & 1.000   & 0.004  & 0.222 & 0.000 & 0.953 \\
                     
                     & $\Xi(0.25)$      &   Firpo  & -     & -   & 1.454  & 0.198 & 2.115 & 0.000     \\
                     &  & CDF    & 0.998 & 1.000   & 0.007  & 0.204 & 0.000 & 0.958 \\
                     
                     & $\Xi(0.5)$           
                     &  Firpo  & -     & -   & 1.475  & 0.179 & 2.177 & 0.000     \\
                     & & CDF    & 0.998 & 1.000   & -0.008 & 0.282 & 0.000 & 0.981 \\
                    
                     & $\Xi(0.75)$        
                     & Firpo  & -     & -   & 1.509  & 0.196 & 2.279 & 0.000     \\
                     &  & CDF    & 0.998 & 1.000   & -0.023 & 0.484 & 0.000 & 0.967 \\

\multicolumn{1}{l}{} & $\Xi(0.8)$   &Firpo  & -     & -   & 1.525  & 0.213 & 2.325 & 0.000     \\
\multicolumn{1}{l}{} &  & CDF    & 0.998 & 1.000   & -0.017 & 0.533 & 0.000 & 0.959 \\ \hline

$1000$               & $\Xi(0.2)$    & Firpo  & -     & -   & 1.447  & 0.153 & 2.096 & 0.000     \\
                     & & CDF    & 0.948 & 1.000   & 0.005  & 0.150 & 0.000 & 0.946 \\

                     & $\Xi(0.25)$     & Firpo  & -     & -   & 1.443  & 0.144 & 2.082 & 0.000     \\
                     &  & CDF    & 0.999 & 1.000   & 0.001  & 0.140 & 0.000 & 0.950 \\

                     & $\Xi(0.5)$     & Firpo  & -     & -   & 1.466  & 0.133 & 2.151 & 0.000     \\
                     & & CDF    & 0.999 & 1.000   & -0.003 & 0.163 & 0.000 & 0.953 \\

                     & $\Xi(0.75)$ & Firpo  & -     & -   & 1.495  & 0.145 & 2.236 & 0.000     \\
                     & & CDF    & 0.999 & 1.000   & -0.007 & 0.298 & 0.000 & 0.955 \\

\multicolumn{1}{l}{} & $\Xi(0.8)$    & Firpo  & -     & -   & 1.502  & 0.147 & 2.257 & 0.000     \\
\multicolumn{1}{l}{} &  & CDF    & 0.999 & 1.000   & -0.007 & 0.350 & 0.000 & 0.956 \\ \hline
 
$2000 $              & $\Xi(0.2)$         & Firpo  & -     & -   & 1.445  & 0.107 & 2.090 & 0.000     \\
                     &  & CDF    & 0.999 & 1.000   & -0.006 & 0.108 & 0.006 & 0.951 \\
                     & $\Xi(0.25)$  & Firpo  & -     & -   & 1.448  & 0.103 & 2.097 & 0.000     \\
                     & & CDF    & 0.999 & 1.000   & -0.006 & 0.100 & 0.003 & 0.944 \\
                     & $\Xi(0.5)$       & Firpo  & -     & -   & 1.467  & 0.090 & 2.153 & 0.000     \\
                     &  & CDF    & 0.999 & 1.000   & -0.007 & 0.125 & 0.004 & 0.954 \\

                     & $\Xi(0.75)$   & Firpo  & -     & -   & 1.502  & 0.099 & 2.258 & 0.000     \\
                     &  & CDF    & 0.999 & 1.000   & -0.010 & 0.237 & 0.037 & 0.967 \\

\multicolumn{1}{l}{} & $\Xi(0.8)$& Firpo  & -     & -   & 1.514  & 0.106 & 2.295 & 0.000     \\
\multicolumn{1}{l}{} &  & CDF    & 0.999 & 1.000   & -0.017 & 0.291 & 0.100 & 0.962 \\ \hline

$10000$              & $\Xi(0.2)$ & Firpo  & -     & -   & 1.443  & 0.047 & 2.084 & 0.000     \\
                     &  & CDF    & 0.971 & 1.000   & -0.002 & 0.047 & 0.000 & 0.949 \\

                     & $\Xi(0.25)$  & Firpo  & -     & -   & 1.445  & 0.045 & 2.089 & 0.000     \\
                     & & CDF    & 0.971 & 1.000   & -0.001 & 0.044 & 0.000 & 0.948 \\

                     & $\Xi(0.5)$ & Firpo  & -     & -   & 1.465  & 0.042 & 2.147 & 0.000     \\
                     & & CDF    & 0.971 & 1.000   & -0.001 & 0.050 & 0.001 & 0.956 \\

                     & $\Xi(0.75)$  & Firpo  & -     & -   & 1.499  & 0.045 & 2.247 & 0.000     \\
                     &  & CDF    & 0.971 & 1.000   & -0.001 & 0.087 & 0.000 & 0.950 \\

\multicolumn{1}{l}{} & $\Xi(0.8)$ & Firpo  & -     & -   & 1.509  & 0.047 & 2.279 & 0.000     \\
\multicolumn{1}{l}{} & & CDF    & 0.971 & 1.000   & -0.000 & 0.103 & 0.000 & 0.947 \\ \hline
\end{longtable}
\end{landscape}

\begin{landscape}
    \begin{longtable}{cccccccc}
    \caption{Simulation results for DTE under Scenario 1}
    \label{DTE-simulation-sc1}\\
    \hline
$n$   &       Estimand        & SEN   & SPE & BIAS   & S.E.  & MSE   & CR    \\
 
    $500$      & $\Delta(0)$  & 0.998 & 1.000   & -0.000 & 0.037 & 0.000 & 0.954 \\
          & $\Delta(3)$  & 0.998 & 1.000   & -0.004 & 0.070 & 0.000 & 0.962 \\
          & $\Delta(-3)$ & 0.998 & 1.000   & 0.149  & 0.008 & 0.022 & 0.961     \\
\hline
 $1000$         & $\Delta(0)$  & 0.999 & 1.000   & -0.000 & 0.028 & 0.000 & 0.958 \\
          & $\Delta(3)$  & 0.999 & 1.000   & 0.000  & 0.054 & 0.000 & 0.959 \\
          & $\Delta(-3)$ & 0.999 & 1.000   & 0.148  & 0.006 & 0.022 & 0.957     \\
\hline
        
    $2000$      & $\Delta(0)$  & 0.999 & 1.000   & -0.000 & 0.019 & 0.000 & 0.948 \\
          & $\Delta(3)$  & 0.999 & 1.000   & -0.000 & 0.037 & 0.000 & 0.959 \\
          & $\Delta(-3)$ & 0.999 & 1.000   & 0.149  & 0.005 & 0.022 & 0.955     \\
    \hline      
 $10000$         & $\Delta(0)$  & 1.000     & 1.000   & -0.000 & 0.008 & 0.000 & 0.952 \\
          & $\Delta(3)$  & 1.000     & 1.000   & -0.000 & 0.016 & 0.000 & 0.953 \\
          & $\Delta(-3)$ & 1.000     & 1.000   & 0.149  & 0.002 & 0.022 & 0.959  \\
          \hline
    \end{longtable}
\end{landscape}

\begin{landscape}

\setlength\tabcolsep{2.5pt}
\footnotesize
    \begin{longtable}{clcccccccccccc}
    \caption{Simulation results for ATE under Scenario 2}
    \label{ATE-simulation}\\
     \hline
      &        & \multicolumn{6}{c}{Hub}                       & \multicolumn{6}{c}{Lattice}   \\  \cmidrule(lr){3-8}\cmidrule(lr){9-14}
   $n$   & Method & SEN   & SPE & BIAS   & S.E.  & MSE    & CR    & SEN   & SPE & BIAS   & S.E.  & MSE    & CR    \\
500  & IPW    & -     & -   & 0.446  & 3.181 & 10.312 & 0.981 & -     & -   & 0.227  & 3.206 & 10.320 & 0.992 \\
      & LD     & -     & -   & 0.929  & 1.047 & 1.960  & 0.924 & -     & -   & 0.530  & 0.869 & 1.036  & 0.969 \\
      & CDF    & 0.911     & 1.000   & 0.929  & 1.047 & 1.960  & 0.924 & 0.961     & 1.000  & 0.530  & 0.869 & 1.036  & 0.969 \\
      \hline
            
 1000   & IPW    & -     & -   & -0.018 & 6.771 & 45.808 & 0.992 & -     & -   & -0.045 & 5.035 & 25.332 & 0.990 \\
      & LD     & -     & -   & 0.678  & 1.130 & 1.736  & 0.969 & -     & -   & 0.391  & 0.899 & 0.961  & 0.971 \\
      & CDF    & 0.945     & 1.000   & 0.678  & 1.130 & 1.736  & 0.969 &  0.982      & 1.000   & 0.391  & 0.899 & 0.961  & 0.971 \\
      \hline

 $ 2000 $   &  IPW    & -     & -   & 0.130  & 5.539 & 30.668 & 0.992 & -     & -   & 0.175  & 1.719 & 2.983  & 0.983 \\
      & LD     & -     & -   & 0.576  & 1.030 & 1.392  & 0.980 & -     & -   & 0.321  & 0.714 & 0.612  & 0.970 \\
      & CDF    & 0.970      & 1.000   & 0.576  & 1.030 & 1.392  & 0.980 & 0.992   & 1.000   & 0.321  & 0.714 & 0.612  & 0.970 \\
      \hline

  $10000$    & IPW    & -     & -   & 0.147  & 2.212 & 4.913  & 0.987 & -     & -   & 0.019  & 2.834 & 8.026  & 0.993 \\
      & LD     & -     & -   & 0.335  & 0.855 & 0.843  & 0.965 & -     & -   & 0.181  & 0.697 & 0.518  & 0.972 \\
      & CDF    &  0.996     & 1.000   & 0.335  & 0.855 & 0.843  & 0.965 & 0.999     & 1.000  & 0.181  & 0.697 & 0.518  & 0.972  \\
      \hline
\end{longtable}
\end{landscape}

\begin{landscape}
    \renewcommand{\arraystretch}{0.88}
    \begin{longtable}{cclcccccccccccc}
    \caption{Simulation results for QTE under Scenario 2}
    \label{QTE-simulation}\\
    \hline
   & && \multicolumn{6}{c}{Hub}   & \multicolumn{6}{c}{Lattice}                       \\ 
   \cmidrule(lr){4-9}\cmidrule(lr){10-15}
$n$                    & \multicolumn{1}{l}{Estimand} & Method & SEN   & SPE & BIAS   & S.E.  & MSE    & CR    & SEN   & SPE & BIAS   & S.E.  & MSE    & CR    \\
$500$                  & \multicolumn{1}{l}{$\Xi(0.2)$}       & Firpo  & -     & -   & 4.121  & 0.375 & 16.987 & 0.000     & -     & -   & 2.813  & 0.285 & 7.913  & 0.000     \\
                     &                             & CDF    & 0.970    & 1.000   & 0.353  & 1.271 & 0.124  & 0.959 & 0.961     & 1.000   & 0.053  & 1.090 & 0.002  & 0.964 \\

                     & \multicolumn{1}{l}{$\Xi(0.25)$}     &Firpo  & -     & -   & 3.934  & 0.338 & 15.482 & 0.000     & -     & -   & 2.751  & 0.261 & 7.572  & 0.000     \\
                     &                           & CDF    & 0.970     & 1.000   & 0.263  & 1.276 & 0.069  & 0.961 & 0.961      & 1.000   & 0.035  & 1.040 & 0.001  & 0.966 \\
     
                     & \multicolumn{1}{l}{$\Xi(0.5)$}      &Firpo  & -     & -   & 3.602  & 0.288 & 12.979 & 0.000     & -     & -   & 2.749  & 0.241 & 7.558  & 0.000     \\
                     &                           & CDF    & 0.970     & 1.000   & 0.156  & 1.284 & 0.024  & 0.951 &  0.961    & 1.000   & 0.009  & 1.066 & 0.000  & 0.959 \\

                     & \multicolumn{1}{l}{$\Xi(0.75)$}     &  Firpo  & -     & -   & 3.973  & 0.321 & 15.786 & 0.000     & -     & -   & 3.277  & 0.295 & 10.742 & 0.000     \\
                     &                            & CDF    &0.970     & 1.000   & 0.460  & 1.343 & 0.211  & 0.951 &  0.961      &1.000  & 0.215  & 1.275 & 0.046  & 0.947 \\

\multicolumn{1}{l}{} & \multicolumn{1}{l}{$\Xi(0.8)$}      & Firpo  & -     & -   & 4.189  & 0.347 & 17.554 & 0.000     & -     & -   & 3.513  & 0.325 & 12.343 & 0.000     \\
\multicolumn{1}{l}{} &                            & CDF    & 0.970    & 1.000   & 0.693  & 1.337 & 0.480  & 0.954 &  0.961      & 1.000  & 0.387  & 1.305 & 0.150  & 0.953 \\ \hline

$1000$                 & \multicolumn{1}{l}{$\Xi(0.2)$}      &                           Firpo  & -     & -   & 4.107  & 0.276 & 16.870 & 0.000     & -     & -   & 2.810  & 0.198 & 7.901  & 0.000     \\
                     &                             & CDF    & 0.948    & 1.000  & 0.243  & 1.177 & 0.059  & 0.957 & 0.992  & 1.000 & -0.036 & 0.993 & 0.001  & 0.962 \\

                     & \multicolumn{1}{l}{$\Xi(0.25)$}   & Firpo  & -     & -   & 3.918  & 0.246 & 15.356 & 0.000     & -     & -   & 2.751  & 0.189 & 7.568  & 0.000     \\
                     &                             & CDF    & 0.948    & 1.000   & 0.197  & 1.114 & 0.039  & 0.961 & 0.992     & 1.000 & -0.040 & 0.971 & 0.001  & 0.967 \\
                     & \multicolumn{1}{l}{$\Xi(0.5)$}     & Firpo  & -     & -   & 3.612  & 0.195 & 13.046 & 0.000     & -     & -   & 2.751  & 0.174 & 7.568  & 0.000     \\
                     &                       & CDF    & 0.948     & 1.000  & 0.178  & 1.060 & 0.032  & 0.971 &0.992 &1.000  & -0.005 & 1.002 & 0.000  & 0.967 \\

                     & \multicolumn{1}{l}{$\Xi(0.75)$}   & Firpo  & -     & -   & 4.001  & 0.232 & 16.008 & 0.000     & -     & -   & 3.279  & 0.210 & 10.751 & 0.000     \\
                     &                         & CDF    & 0.948    &1.000  & 1.397  & 1.237 & 0.158  & 0.956 & 0.992   & 1.000  & 0.154  & 1.176 & 0.023  & 0.945 \\

\multicolumn{1}{l}{} & \multicolumn{1}{l}{$\Xi(0.8)$}      & Firpo  & -     & -   & 4.214  & 0.255 & 17.759 & 0.000     & -     & -   & 3.512  & 0.228 & 12.335 & 0.000     \\
\multicolumn{1}{l}{} &           & CDF    & 0.948   & 1.000   & 0.541  & 1.289 & 0.293  & 0.967 & 0.992   & 1.000   & 0.292  & 1.184 & 0.085  & 0.951 \\

\hline
 
$2000 $                & \multicolumn{1}{l}{$\Xi(0.2)$}& Firpo  & -     & -   & 4.113  & 0.186 & 16.924 & 0.000     & -     & -   & 2.823  & 0.140 & 7.971  & 0.000     \\
                     &                          & CDF    &  0.971     & 1.000  & 0.082  & 1.362 & 0.006  & 0.959 &  0.981    &1.000   & 0.013  & 0.769 & 0.000  & 0.980 \\

                     & \multicolumn{1}{l}{$\Xi(0.25)$}    & Firpo  & -     & -   & 3.928  & 0.170 & 15.431 & 0.000     & -     & -   & 2.762  & 0.134 & 7.628  & 0.000     \\
                     &   & CDF    & 0.971     &1.000 & 0.056  & 1.312 & 0.003  & 0.965 & 0.981   & 1.000  & 0.021  & 0.971 & 0.000  & 0.983 \\
                  
                     & \multicolumn{1}{l}{$\Xi(0.5)$}   & Firpo  & -     & -   & 3.607  & 0.135 & 13.016 & 0.000     & -     & -   & 2.760  & 0.115 & 7.621  & 0.000     \\
                     &  & CDF    &  0.971  & 1.000   & 0.063  & 1.130 & 0.004  & 0.967 & 0.981    & 1.000   & 0.047  & 1.002 & 0.002  & 0.982 \\

                     & \multicolumn{1}{l}{$\Xi(0.75)$}  & Firpo  & -     & -   & 3.997  & 0.155 & 15.981 & 0.000     & -     & -   & 3.277  & 0.140 & 10.742 & 0.000     \\
                     & & CDF    &  0.971    &1.000   & 0.193  & 1.265 & 0.037  & 0.945 & 0.981     & 1.000  & 0.085  & 1.176 & 0.007  & 0.953 \\

\multicolumn{1}{l}{} & \multicolumn{1}{l}{$\Xi(0.8)$}     & Firpo  & -     & -   & 4.219  & 0.178 & 17.806 & 0.000     & -     & -   & 3.509  & 0.156 & 12.314 & 0.000     \\
\multicolumn{1}{l}{} &        & CDF    &  0.971     & 1.000   & 0.317  & 1.292 & 0.100  & 0.946 & 0.981     & 1.000   & 0.160  & 1.184 & 0.025  & 0.943 \\
\hline

$10000$                & \multicolumn{1}{l}{$\Xi(0.2)$}    & Firpo  & -     & -   & 4.110  & 0.085 & 16.893 & 0.000     & -     & -   & 2.817  & 0.065 & 7.935  & 0.000     \\
                     &                          & CDF    & 0.971 & 1.000  & -0.015 & 1.013 & 0.000  & 0.972 &  0.981  & 1.000 & 0.010  & 0.484 & 0.000  & 0.985 \\

                     & \multicolumn{1}{l}{$\Xi(0.25)$}  & Firpo  & -     & -   & 3.921  & 0.076 & 15.377 & 0.000     & -     & -   & 2.755  & 0.060 & 7.593  & 0.000     \\
                     &                          & CDF    & 0.971  & 1.000 & 0.003  & 0.896 & 0.000  & 0.977 & 0.981     & 1.000   & 0.056  & 1.312 & 0.000  & 0.985 \\

                     & \multicolumn{1}{l}{$\Xi(0.5)$}   & Firpo  & -     & -   & 3.604  & 0.062 & 12.990 & 0.000     & -     & -   & 2.757  & 0.055 & 7.604  & 0.000     \\
                     &   & CDF    & 0.971  & 1.000 & 0.044  & 0.719 & 0.001  & 0.979 & 0.981   & 1.000  & 0.063  & 1.130 & 0.000  & 0.991 \\

                     & \multicolumn{1}{l}{$\Xi(0.75)$}    & Firpo  & -     & -   & 3.989  & 0.071 & 15.919 & 0.000     & -     & -   & 3.276  & 0.064 & 10.736 & 0.000     \\
                     &     & CDF    &0.971   & 1.000  & 0.020  & 1.205 & 0.000  & 0.961 & 0.981   &1.000 & 0.193  & 1.265 & 0.000  & 0.967 \\

\multicolumn{1}{l}{} & \multicolumn{1}{l}{$\Xi(0.8)$}      & Firpo  & -     & -   & 4.210  & 0.079 & 17.729 & 0.000     & -     & -   & 3.509  & 0.070 & 12.316 & 0.000     \\
\multicolumn{1}{l}{} &                 & CDF    & 0.971   & 1.000  & 0.056  & 1.318 & 0.000  & 0.953 & 0.981  &1.000  & 0.317  & 1.292 & 0.000  & 0.958\\
\hline
    \end{longtable}
    
\end{landscape}
\begin{landscape}
    \begin{center}
        \begin{longtable}{cccccccccccccc}
        
        \caption{Simulation results for DTE under Scenario 2}
        \label{DTE-simulation}\\
    \hline
        &          & \multicolumn{6}{c}{Hub}                      & \multicolumn{6}{c}{Lattice}                  \\ \cmidrule(lr){3-8}\cmidrule(lr){9-14}
$n$   &   Estimand       & SEN   & SPE & BIAS   & S.E.  & MSE   & CR    & SEN   & SPE & BIAS   & S.E.  & MSE   & CR    \\
   $500$     &  $\Delta(0)$        & 0.913    & 1.000   & -0.066 & 0.103 & 0.004 & 0.928 & 0.961     & 1.000   & -0.039 & 0.079 & 0.001 & 0.939 \\
        & $\Delta(3)$         & 0.913    & 1.000  & -0.088 & 0.150 & 0.007 & 0.957 & 0.961  &1.000 & -0.058 & 0.134 & 0.003 & 0.962 \\
        & $\Delta(-3)$        &  0.913   & 1.000   & -0.044 & 0.084 & 0.002 & 0.975 & 0.961   & 1.000 & -0.007 & 0.056 & 0.000 & 0.971 \\
        \hline
$1000$  &   $\Delta(0)$         & 0.945   & 1.000  & -0.045 & 0.091 & 0.002 & 0.964 & 0.982    & 1.000   & -0.024 & 0.071 & 0.000 & 0.959 \\
        & $\Delta(3)$         &0.945  &1.000  & -0.052 & 0.137 & 0.002 & 0.960 & 0.982 &1.000   & -0.041 & 0.117 & 0.001 & 0.963 \\
        & $\Delta(-3)$        & 0.945  & 1.000  & -0.026 & 0.096 & 0.000 & 0.959 &0.982    & 1.000  & 0.001  & 0.050 & 0.000 & 0.950 \\
        \hline
$2000$  &  $\Delta(0)$         &0.970 &1.000   & -0.031 & 0.079 & 0.000 & 0.966 &0.992  & 1.000  & -0.016 & 0.058 & 0.000 & 0.956 \\
        & $\Delta(3)$         &0.970   & 1.000   & -0.044 & 0.109 & 0.001 & 0.964 &0.992  & 1.000  & -0.029 & 0.103 & 0.000 & 0.961 \\
        & $\Delta(-3)$        & 0.970 & 1.000  & -0.022 & 0.078 & 0.000 & 0.967 &0.992 & 1.000  & 0.004  & 0.041 & 0.000 & 0.961 \\
        \hline
$10000$ &  $\Delta(0)$    & 0.996    & 1.000   & -0.014 & 0.074 & 0.000 & 0.964 &0.999 & 1.000   & -0.008 & 0.036 & 0.000 & 0.963 \\
        & $\Delta(3)$      &0.996   & 1.000   & -0.023 & 0.074 & 0.000 & 0.973 &0.999  & 1.000   & -0.010 & 0.071 & 0.000 & 0.962 \\
        & $\Delta(-3)$     & 0.996 &1.000  & -0.009 & 0.061 & 0.000 & 0.972 &0.999  & 1.000   & 0.009  & 0.028 & 0.000 & 0.946\\
        \hline
\end{longtable}
    \end{center}
\end{landscape}

\renewcommand{\arraystretch}{0.85}
\begin{longtable}[h!]{llrcl}
\caption{Real data analysis: estimation of various treatment effcts by the proposed and existing methods.}\\
    \hline
    Estimand & Method    & EST  & S.E. & p-value  \\
    \hline
$\text{ATE} $     
         & IPW    & 3.841                      & 1.128                     & 0.000\\
         & LD     & 3.568                     & 0.894                     & 0.000 \\
         & CDF    & 3.568                     & 0.894                    & 0.000 \\
         \hline
$\Xi(0.2)$  
         & Firpo  & 3.430                     & 1.107                     & 0.000 \\
         & CDF    & 2.686                      & 0.864                     & 0.000\\
         \hline
$\Xi(0.25)$  
         & Firpo  & 3.933                    & 1.344                     & 0.001 \\
         & CDF    & 2.803                     & 1.302                     & 0.015\\
         \hline
$\Xi(0.5)$ 

         & Firpo  & 4.618                     & 1.500                     & 0.001 \\
         & CDF    & 3.284                     & 1.342                     & 0.007\\
         \hline
$\Xi(0.75)$  
         & Firpo  & 3.263                      & 1.521                    & 0.015 \\
         & CDF    & 2.403                  & 1.514                     & 0.056\\
         \hline
$\Xi(0.8)$   
         & Firpo  & 3.270                      & 1.409                    & 0.010\\
         & CDF    & 2.476                     & 1.259                    & 0.024 \\
         \hline
$\Delta(\Bar{Y})$    
         & CDF    & \multicolumn{1}{l}{-0.072} & \multicolumn{1}{l}{0.027} & 0.004 \\
         \hline

    \label{estimated causal effects}
\end{longtable}

\begin{figure}[!ht]
\begin{center}
\begin{tikzpicture}
% nodes %
\node at ( -2,-2)[rectangle,draw] (A) {$A$};
\node at ( 2,-2) (Y)[rectangle,draw] {$Y$};
%%%%%%%%
\node at ( -1,0) (x3)[circle,draw] {$X_3$};
\node at ( 1,0) (x1)[circle,draw] {$X_1$};
\node at ( -1,-4) (x4)[circle,draw] {$X_4$};
\node at ( 1,-4) (x5)[circle,draw] {$X_5$};
\node at ( 2.5,2) (x2)[circle,draw] {$X_2$};
%edge
\draw[->, line width= 1] (x3) --  (A);
\draw[->, line width= 1] (x4) --  (A);
\draw[->, line width= 1] (x4) --  (Y);
\draw[->, line width= 1] (x1) --  (A);
\draw [->, line width= 1] (x1) -- (Y);
\draw [->, line width= 1] (x3) -- (Y);
\draw[->, line width= 1] (A) --  (Y);
\draw [->,  line width= 1] (x2) -- (Y);
\draw [dashed, color=red,  line width= 1] (x2) -- (x1);
\draw [dashed, color=red,  line width= 1] (x3) -- (x1);
\draw [dashed, color=red,  line width= 1] (x2) -- (x3);
\end{tikzpicture}
\end{center}
\caption{Diagram of causal inference with network covariates} \label{Causal-Network-fig}
\end{figure}
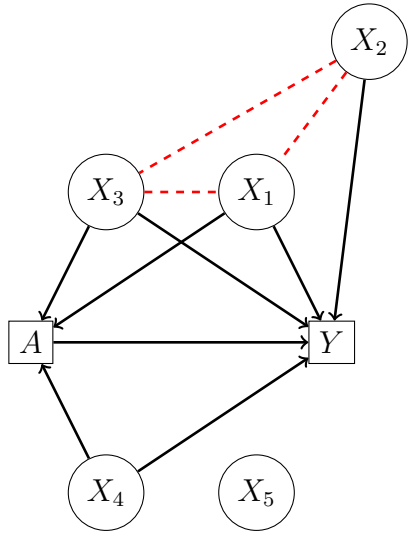
}

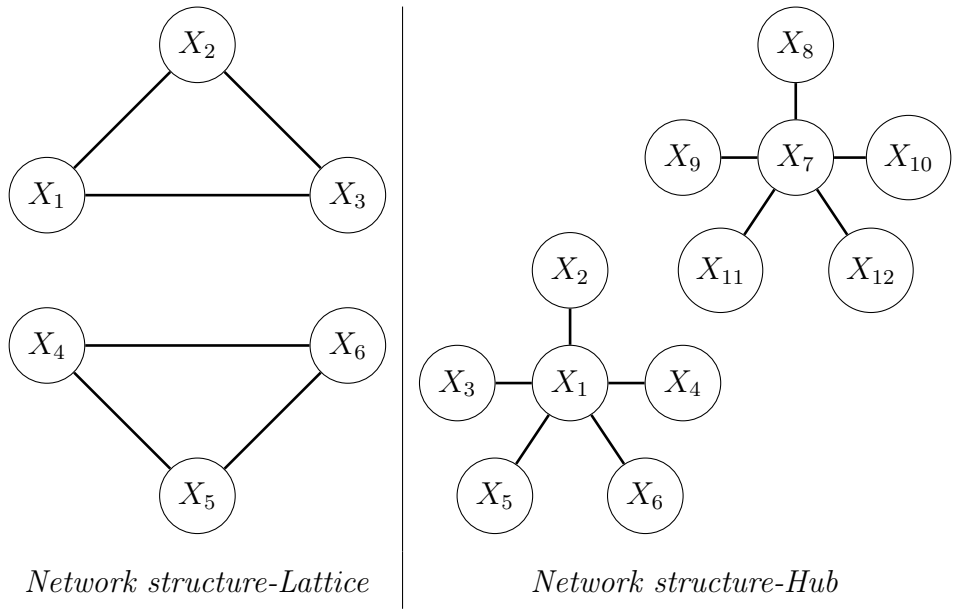
\begin{figure}[!ht]
\begin{center}
\begin{tabular}{c | c}
 \begin{tikzpicture}
% nodes %

\node at ( -2,0)[circle,draw] (1) {$X_1$};
\node at ( 0,2)[circle,draw] (2) {$X_2$};
\node at ( 2,0)[circle,draw] (3) {$X_{3}$};

\node at ( -2,-2)[circle,draw] (4) {$X_4$};

\node at ( 0,-4)[circle,draw] (5) {$X_5$};
\node at ( 2,-2)[circle,draw] (6) {$X_{6}$};

%edge
\draw[-, line width= 1] (1) --  (2);
\draw[-, line width= 1] (2) --  (3);
\draw[-, line width= 1] (3) --  (1);
\draw[-, line width= 1] (4) --  (5);
\draw[-, line width= 1] (5) --  (6);
\draw[-, line width= 1] (6) --  (4);

\end{tikzpicture} &
\begin{tikzpicture}
% nodes %
\node at ( 1.5,0)[circle,draw] (4) {$X_4$};
\node at ( -1.5,0)[circle,draw] (3) {$X_3$};
\node at ( -1.0,-1.5)[circle,draw] (5) {$X_5$};
\node at ( 0,1.5)[circle,draw] (2) {$X_2$};
\node at ( 0,0)[circle,draw] (1) {$X_1$};
\node at ( 1.0,-1.5)[circle,draw] (6) {$X_6$};
%%%%%%%%%%
\node at ( 4.5,3)[circle,draw] (10) {$X_{10}$};
\node at ( 1.5,3)[circle,draw] (9) {$X_9$};
\node at ( 2.0,1.5)[circle,draw] (11) {$X_{11}$};
\node at ( 3,4.5)[circle,draw] (8) {$X_8$};
\node at ( 3,3)[circle,draw] (7) {$X_7$};
\node at ( 4.0,1.5)[circle,draw] (12) {$X_{12}$};

%edge
\draw[-, line width= 1] (1) --  (2);
\draw[-, line width= 1] (1) --  (3);
\draw[-, line width= 1] (1) --  (4);
\draw[-, line width= 1] (1) --  (5);
\draw[-, line width= 1] (1) --  (6);
%%%
\draw[-, line width= 1] (7) --  (8);
\draw[-, line width= 1] (7) --  (9);
\draw[-, line width= 1] (7) --  (10);
\draw[-, line width= 1] (7) --  (11);
\draw[-, line width= 1] (7) --  (12);
\end{tikzpicture}
 \\
  \textit{Network structure-Lattice} & \textit{Network structure-Hub}  
\end{tabular}
\end{center} 
\caption{Covariate network structures for simulation studies. The left panel displays the lattice structure, where $X_1$, $X_2$ and $X_3$ form a triangle, $X_4$, $X_5$ and $X_6$ form the other, and $X_7$ to $X_{12}$ are not connected; the right panel displays the hub structure, which $X_1$ to $X_6$ and $X_7$ to $X_{12}$ are interwoven, respectively.}\label{Sim-Network-fig}
\end{figure}

\begin{landscape}
    \begin{figure}
        \centering
        \includegraphics[scale = 0.9]{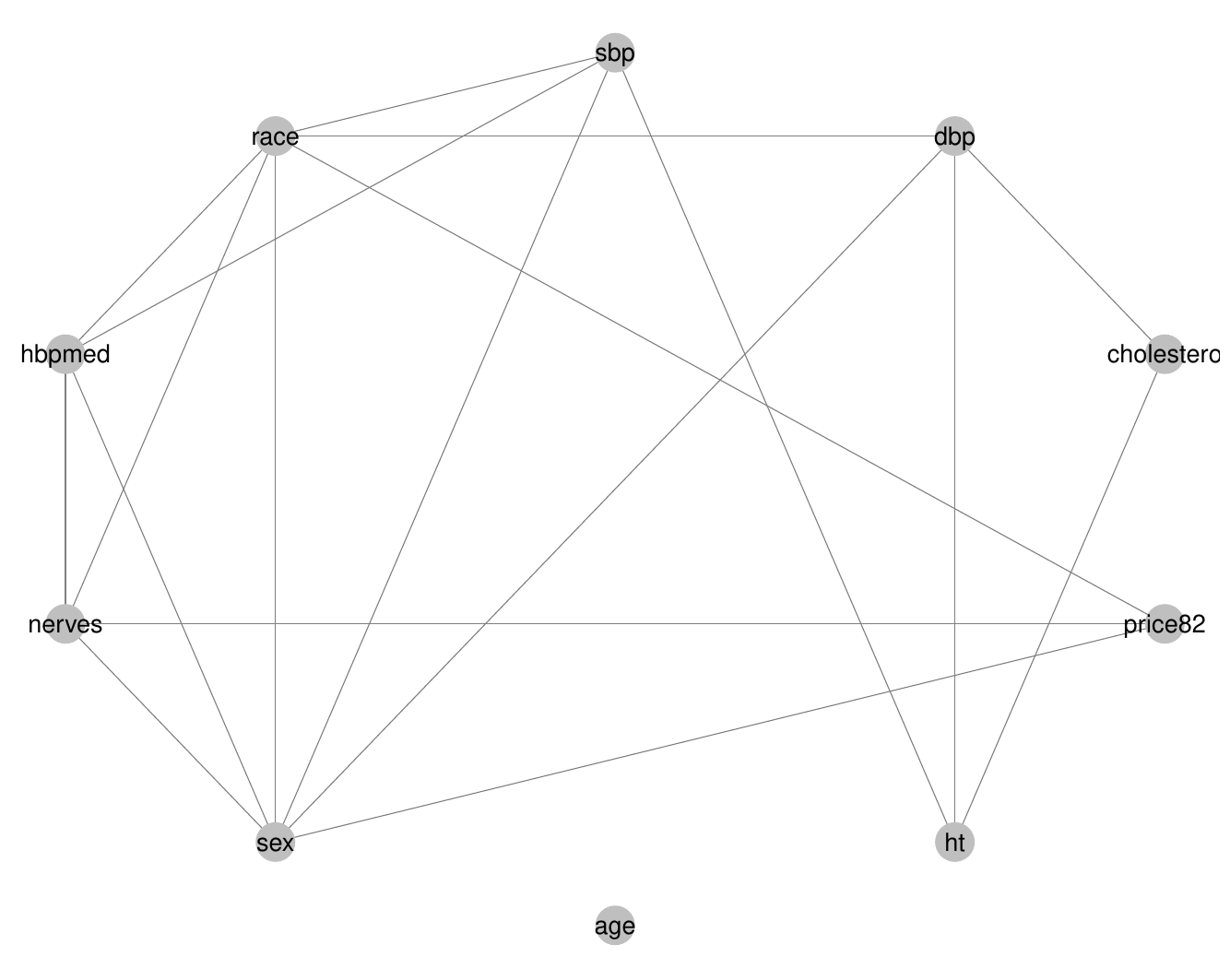}
        \caption{Real data analysis result: estimated network structure for nine confounders in NHEFS data.}
        \label{0.01}
    \end{figure}
    
\end{landscape}

\thispagestyle{empty}
\setcounter{footnote}{0}
\setcounter{section}{0}
\newtheorem{lemma}[theorem]{Lemma}
\numberwithin{equation}{section}
\renewcommand{\thesection}{\alph{section}}

\hfill March 27, 2025\ \\

\baselineskip=28pt
\begin{center}
{\LARGE{\bf Supporting Information for “A Unified Approach for Estimating Various Treatment Effect in Causal Inference”}}
\end{center}
\baselineskip=14pt
\vskip 2mm
\begin{center}
Kuan-Hsun Wu and Li-Pang Chen\footnote{Correpsonding Author. Email: \hyperlink{mailto:lchen723@nccu.edu.tw}{lchen723@nccu.edu.tw}}
\\~\\

\textit{Department of Statistics, National Chengchi University}
\end{center}
\bigskip

\begin{center}
{\Large{\bf Abstract}}
\end{center}
\baselineskip=17pt
{

The supporting information contains a list of regularity conditions and the detailed proofs of theorems in the manuscript, entitled “A Unified Approach for Estimating Various Treatment Effect in Causal Inference”.
}

\par\vfill\noindent
\underline{\bf Keywords}: Causal inference; counterfactual distributions; plug-in estimators; high-dimensional;  network structures; graphical model; propensity score; variable selection.

\par\medskip\noindent
\underline{\bf Short title}: Estimating Various Cauasl Effects Using IPW CDF

\clearpage\pagebreak\newpage
\pagenumbering{arabic}

\setlength{\gnat}{22pt}
\baselineskip=\gnat

\clearpage
\setcounter{footnote}{0}
\setcounter{section}{0}
\appendix
{\section{{Regularity Conditions}}}
We list the following conditions that are imposed to the derivation:
 \begin{itemize}
    \item[(A1)] {The matrices $\bold{X}_{V}^\top \bold{X}_{V}$ and $(\bold{X}^{\circ2}_{E})^\top\bold{X}^{\circ2}_{E}$ are nonsingular, where $\bold{X}_V$ and $\bold{X}^{\circ2}_E$ are the matrices consisting of only the variables whose indices are recorded by sets $V$ and $E$, respectively. That is, the matrices $(\bold{X}_{V}^\top \bold{X}_{V})^{-1}$ and $[(\bold{X}^{\circ2}_{E})^\top\bold{X}^{\circ2}_{E}]^{-1}$ exist.}
     \item[(A2)] There exist two constants $\gamma_1$ and $\gamma_2\in(0,1)$, such that
     $$
    \underset{i\in V^C}{\text{max}} ||(\bold{X}_V^\top \bold{X}_V)^{-1}\bold{X}^\top_V\bold{X}_i||_1\leq 1-\gamma_1\text{ and }\underset{i\in E^C}{\text{max}} ||[(\bold{X}^{\circ2}_E)^\top (\bold{X}^{\circ2}_E)]^{-1}(\bold{X}^{\circ2}_E)^\top\bold{X}^{\circ2}_i||_1\leq 1-\gamma_2,
    $$
    where $\bold{X}_i$ is the $i$th column of matrix $\bold{X}$ and $\bold{X}^{\circ2}_i$ is the $i$th column of matrix $\bold{X}^{\circ2}$.
     
     \item[(A3)] There exist four positive constants $a_1$, $a_2$, $b_1$ and $b_2$, such that
    $$
            E(\mathbb{X}_r) < a_1, E(\mathbb{X}_r^{\top}{\mathbb{X}_r}) < a_2, E(\mathbb{X}^{\circ2}_r) < b_1 \text{ and } E\bigg\{(\mathbb{X}^{\circ2})^{\top}{(\mathbb{X}^{\circ2})}\bigg\} < b_2
    $$
    where $\mathbb{X}_r$ is the $r$th column of matrix $\mathbb{X}$.
    
     \item[(A4)] 
     There exist four positive constants $c_1$, $c_2$, $d_1$ and $d_2$, such that
    $$
    \boldsymbol{\Lambda}_{\text{min}}(\mathbb{X}) > c_1, \boldsymbol{\Lambda}_{\text{max}}(\mathbb{X}) < c_2, \boldsymbol{\Lambda}_{\text{min}}(\mathbb{X}^{\circ2}) > d_1\text{ and }\boldsymbol{\Lambda}_{\text{max}}(\mathbb{X}^{\circ2}) < d_2.
    $$
    where $\boldsymbol{\Lambda}_{\text{min}}(\bold{A})$ and $\boldsymbol{\Lambda}_{\text{max}}(\bold{A})$ represent the minimum and maximum eigenvalues of the matrix $\bold{A}$, respectively.
    \item[(A5)] There exists a positive constant $K_{\text{clm}}$, such that
    $$
    \underset{j = 1,2,\dots,p}{\text{max}} \frac{||\bold{X}_j||_2}{\sqrt{n}} \leq K_{\text{clm}}.
    $$
    \item[(A6)] The true CDFs of $Y^{(1)}$ and  $Y^{(0)}$ are continuous. Moreover, the derivatives of $F^{(1)}$ and $F^{(0)}$ exist, and the corresponding density functions are denoted by $f^{(1)}(y)$ and $f^{(0)}(y)$ of $Y^{(1)}$ and $Y^{(0)}$, respectively.
 \end{itemize}
Assumptions (A1)-(A{5}) state the conditions on the behavior of design matrices $\bold{X},\bold{X}^{\circ2}, \mathbb{X}$ and $\mathbb{X}^{\circ2}$, ensuring the sparsity recovery for main and interaction effects holds. {Specifically, (A1) is necessary for computing the regression coefficients using $\bold{X}_V$ and $\bold{X}^{\circ2}_E$. (A2) is often called {\em mutual incoherence}, which restricts the dependence of variables between in $V$ (or $E$) and $V^C$ (or $E^C$). (A3) is used to guarantee the existence of variance of confounders. The eigenvalues in (A4) are required to ensure that the covariates are not excessively linear dependent (e.g., Yang et al., 2015). In addition, (A5) states that the columns of the design matrix are normalized. Following the last remark of Section \ref{Sec2.1}, (A6) concentrates our discussion on the continuous outcome.}

\section{{Some Lemmas}}
In this section, we introduce some useful lemmas for the derivations of the main theorems.
\begin{lemma}[Gaussian Tail Bound.]\label{tail bound}
    Let $X\sim\text{N}(\mu,\sigma^2)$. The Gaussian tail bound is, for any $\epsilon>0$,
     $$
     P(|X|>\epsilon)\leq 2\exp{\bigg(\frac{(\epsilon-\mu)^2}{2\sigma^2}\bigg)}.
    $$
   
\end{lemma}
{We demonstrate the derivation of Lemma \ref{tail bound} by using the Chernoff's bound, which states that, for any $\epsilon>0$,}
\begin{equation}\label{chernoff}
    P(X-\mu>\epsilon)\leq \exp{\bigg(\frac{\epsilon^2}{2\sigma^2}\bigg)}.
\end{equation}
    Let $\epsilon^* = \epsilon -\mu$. Directly applying (\ref{chernoff}) yields that
    $$
    \begin{aligned}
        P(X>\epsilon) &= P(X-\mu>\epsilon-\mu)\\
        &=P(X-\mu>\epsilon^*)\\
        &\leq  \exp{\bigg(\frac{{\epsilon^*}^2}{2\sigma^2}\bigg)}\\
        &= \exp{\bigg(\frac{(\epsilon-\mu)^2}{2\sigma^2}\bigg)}.
    \end{aligned}
    $$
{Due to the symmetry of normally distributed random variables and the union bound, the Gaussian tail bound is obtained.}
\begin{lemma}[Expansion of Statistical Functionals.]\label{asymptotic normallity stat fun}
    Assume that $\widehat{F}$, an estimated CDF derived by the IID sample, converges to $F$ in probability {uniformly over $y\in\mathbb{R}$}, where $F$ is a CDF of a random variable $Y$. Consider a statistical functional $T$ with influence {curve} $\phi_F$ {defined in Section \ref{Theory}.  Therefore, the asymptotic variance of $\sqrt{n}\bigg(T(\widehat{F})-T(F)\bigg)$ is determined only by the dominant term $\dfrac{1}{\sqrt{n}}\sum\limits_{i=1}^n\phi_F(Y_i)$.\\
    \underline{\bf{Proof:}}\\
    Let $D[0,1]$ be the space of right-continuous real-valued functions on $[0,1]$ with left-hand limits. That is, for any $R\in D[0,1]$, we have that
    $$
    \lim_{t\to t^+}R(t) = R(t)
    $$
    for any $t\in[0,1)$ and
    $$
    \lim_{t\to t^-}R(t)
    $$
    exists for any $t\in(0,1]$. We define $\rho:D[0,1]\to\mathbb{R}$, the functional induced by a statistical functional $T$, as
    $$
    \rho(R) = T(R\circ F)
    $$
    Let $U$ denote the distribution function from the uniform distribution in an interval $[0,1]$, say
    \begin{equation}\label{uniform distribution}
        U(t) = \begin{cases}
        0&\textup{ if }t<0;\\
        t&\textup{ if }0\leq t<1;\\
        1&\textup{ if }t\ge1.
    \end{cases}
    \end{equation}
    We have that $U = F\circ F^{-1}$ and $\widehat{U} \triangleq \widehat{F}\circ F^{-1}$, where $F^{-1}$ is the inverse distribution function of $F$.  By Lemma 4.4.1 of Fernholz (1983), if the induced functional $\rho$ is Hadamard differentiable at the uniform distribution on $[0,1]$, that is, there exists a continuous linear function $\rho^\prime_U$ such that}
    $$
\frac{\rho(U+t_nh_n) - \rho(U)}{t_n} \to\rho^\prime_U(h)\textup{ as }n\to\infty,
    $$
    for all sequences $\{t_n\}\subset\mathbb{R}$ and $\{h_n\}\in D[0,1]$ satisfying $t_n\to0$, $h_n\to h\in D[0,1]$ and $U+t_nh_n\in D[0,1]$, then combining with (\ref{influence curve}) gives that
\begin{equation}\label{influence curve and Hadamard derivative}
    \phi_F(y) = \rho^\prime_U((\delta_y-U)\circ F^{-1}).
\end{equation}
An application of Hadamard derivative yields that
\begin{equation}\label{expansion}
        \begin{aligned}
            \sqrt{n}\bigg(T(\widehat{F}) - T(F)\bigg) &= \sqrt{n}\bigg(\rho_U^\prime(\widehat{U}-U)\bigg)+\sqrt{n}\textup{Rem}(\widehat{U}-U)\\
    &=\sqrt{n}\Bigg[\rho_U^\prime\bigg((\widehat{F}-{F})\circ F^{-1}\bigg)\Bigg]+\sqrt{n}\textup{Rem}(\widehat{U}-U)\\
       &=\sqrt{n}\Bigg[\rho_U^\prime\bigg(\frac{1}{n}\sum_{i=1}^n(\delta_{Y_i}-{F})\circ F^{-1}\bigg)\Bigg]+\sqrt{n}\textup{Rem}(\widehat{U}-U)\\
       &=\frac{1}{\sqrt{n}}\sum_{i=1}^n\phi_F(Y_i)+\sqrt{n}\textup{Rem}(\widehat{U}-U),
    \end{aligned}
\end{equation}
where the first equality is shown in Fernholz (1983, p.40), last equality is obtained with (\ref{influence curve and Hadamard derivative}) and $\textup{Rem}(\cdot)$ is the remainder term defined in Fernholz (1983). By Proposition 4.3.3 and Theorem 4.4.2 in Fernholz (1983), we have that $\textup{Rem}(\widehat{U}-U)\overset{p}{\to}0$ as $n\to\infty$. Therefore, the claim is justified.
\end{lemma}
{\section{{Justification of CDF of (\ref{est-general treatment effect}) and (\ref{9})}}}\label{justification}
To verify that (\ref{est-general treatment effect}) and (\ref{9}) form CDF, it suffices to check the properties of CDF. Specifically,
\begin{equation} \label{CDF-1-to1}
    \begin{aligned}
        \lim_{y\to\infty} \widehat{F}^{(1)}(y) &= \bigg(\sum_{i=1}^n\dfrac{A_i}{\widehat{\pi}(\bold{X}_i)}\bigg)^{-1}\lim_{y\to\infty} \sum_{i=1}^n\dfrac{A_i I(Y_i\leq y)}{\widehat{\pi}(\bold{X}_i)} \\
         &=\bigg(\sum_{i=1}^n\dfrac{A_i}{\widehat{\pi}(\bold{X}_i)}\bigg)^{-1} \bigg(\sum_{i=1}^n\dfrac{A_i}{\widehat{\pi}(\bold{X}_i)}\bigg)\\
         &=1
    \end{aligned}
\end{equation}
and
\begin{equation}\label{CDF-0-to1}
    \begin{aligned}
        \lim_{y\to\infty} \widehat{F}^{(0)}(y) &= \bigg(\sum_{i=1}^n\dfrac{1-A_i}{1-\widehat{\pi}(\bold{X}_i)}\bigg)^{-1}\lim_{y\to\infty} \sum_{i=1}^n\dfrac{(1-A_i) I(Y_i\leq y)}{1-\widehat{\pi}(\bold{X}_i)} \\
        &=\bigg(\sum_{i=1}^n\dfrac{1-A_i}{1-\widehat{\pi}(\bold{X}_i)}\bigg)^{-1} \bigg(\sum_{i=1}^n\dfrac{(1-A_i)}{1-\widehat{\pi}(\bold{X}_i)}\bigg)\\
        &=1.
    \end{aligned}
\end{equation}
Similarly, taking the limit $y\to-\infty$ to (\ref{est-general treatment effect}) and (\ref{9}) gives that
\begin{equation}\label{CDF-1-to0}
    \begin{aligned}
        \lim_{y\to-\infty} \widehat{F}^{(1)}(y) &= \bigg(\sum_{i=1}^n\dfrac{A_i}{\widehat{\pi}(\bold{X}_i)}\bigg)^{-1}\lim_{y\to-\infty} \sum_{i=1}^n\dfrac{A_i I(Y_i\leq y)}{\widehat{\pi}(\bold{X}_i)} \\
         &=\bigg(\sum_{i=1}^n\dfrac{A_i}{\widehat{\pi}(\bold{X}_i)}\bigg)^{-1} \times0\\
         &=0
    \end{aligned}
\end{equation}
and
\begin{equation}\label{CDF-0-to0}
\begin{aligned}
        \lim_{y\to-\infty} \widehat{F}^{(0)}(y) &= \bigg(\sum_{i=1}^n\dfrac{1-A_i}{1-\widehat{\pi}(\bold{X}_i)}\bigg)^{-1}\lim_{y\to-\infty} \sum_{i=1}^n\dfrac{(1-A_i) I(Y_i\leq y)}{1-\widehat{\pi}(\bold{X}_i)} \\
        &=\bigg(\sum_{i=1}^n\dfrac{1-A_i}{1-\widehat{\pi}(\bold{X}_i)}\bigg)^{-1} \times 0\\
        &=0.
\end{aligned}
\end{equation}
Finally, we check the non-decreasing property of (\ref{est-general treatment effect}) and (\ref{9}). We denote the ordered sample of $\{Y_1,Y_2,\dots,Y_n\}$ as $\{Y_{(1)},Y_{(2)},\dots,Y_{(n)}\}$ with $Y_{(i)}\leq Y_{(j)}$ for any $i<j$. For $i = 1,2,\dots,n-1$, define interval $\bold{I}_i \triangleq(Y_{(i)},Y_{(i+1)}]$ with $\bold{I}_0 \triangleq (-\infty,Y_{(1)}]$ and $\bold{I}_n\triangleq (Y_{(n)},\infty)$. For any $a< b$, let
$$
a^* \triangleq \sup_m \bigg\{m: a\in\bigcup_{i=0}^m\bold{I}_i\bigg\}\text{ and }b^* \triangleq \sup_m \bigg\{m:b\in\bigcup_{i=0}^m\bold{I}_i\bigg\}.
$$
As $\bigg\{m: a\in\bigcup\limits_{i=0}^m\bold{I}_i\bigg\}\subseteq\bigg\{m:b\in\bigcup\limits_{i=0}^m\bold{I}_i\bigg\}$, we have $a^{*}\leq b^{*}$. Therefore,
\begin{equation}\label{F1-monotone}
    \allowdisplaybreaks[1]
\begin{aligned}
    \widehat{F}^{(1)}(b) &= \bigg(\sum_{i=1}^n\frac{A_i}{\widehat{\pi}(\bold{X}_i)}\bigg)^{-1}\sum_{i=1}^n\dfrac{A_i I(Y_i\leq b)}{\widehat{\pi}(\bold{X}_i)}\\
    &= \bigg(\sum_{i=1}^n\frac{A_i}{\widehat{\pi}(\bold{X}_i)}\bigg)^{-1}\sum_{i=1}^n\dfrac{A_i I(Y_{(i)}\leq b)}{\widehat{\pi}(\bold{X}_i)}\\
    &=\bigg(\sum_{i=1}^n\frac{A_i}{\widehat{\pi}(\bold{X}_i)}\bigg)^{-1}\sum_{i=1}^{b^*}\dfrac{A_i I(Y_{(i)}\leq b)}{\widehat{\pi}(\bold{X}_i)}\\
    &\ge \bigg(\sum_{i=1}^n\frac{A_i}{\widehat{\pi}(\bold{X}_i)}\bigg)^{-1}\sum_{i=1}^{a^*}\dfrac{A_i I(Y_{(i)}\leq b)}{\widehat{\pi}(\bold{X}_i)}\\
    &= \bigg(\sum_{i=1}^n\frac{A_i}{\widehat{\pi}(\bold{X}_i)}\bigg)^{-1}\sum_{i=1}^{a^*}\dfrac{A_i I(Y_{(i)}\leq a)}{\widehat{\pi}(\bold{X}_i)}\\
    &=\widehat{F}^{(1)}(a)
\end{aligned}
\end{equation}
and
\begin{equation}\label{F0-monotone}
\begin{aligned}
    \widehat{F}^{(0)}(b) &= \bigg(\sum_{i=1}^n\frac{1-A_i}{1-\widehat{\pi}(\bold{X}_i)}\bigg)^{-1}\sum_{i=1}^n\dfrac{(1-A_i) I(Y_i\leq b)}{1-\widehat{\pi}(\bold{X}_i)}\\
    &= \bigg(\sum_{i=1}^n\frac{1-A_i}{1-\widehat{\pi}(\bold{X}_i)}\bigg)^{-1}\sum_{i=1}^n\dfrac{(1-A_i) I(Y_{(i)}\leq b)}{1-\widehat{\pi}(\bold{X}_i)}\\
    &=\bigg(\sum_{i=1}^n\frac{1-A_i}{1-\widehat{\pi}(\bold{X}_i)}\bigg)^{-1}\sum_{i=1}^{b^*}\dfrac{(1-A_i) I(Y_{(i)}\leq b)}{\widehat{\pi}(\bold{X}_i)}\\
    &\ge \bigg(\sum_{i=1}^n\frac{1-A_i}{1-\widehat{\pi}(\bold{X}_i)}\bigg)^{-1}\sum_{i=1}^{a^*}\dfrac{(1-A_i) I(Y_{(i)}\leq b)}{\widehat{\pi}(\bold{X}_i)}\\
    &= \bigg(\sum_{i=1}^n\frac{1-A_i}{1-\widehat{\pi}(\bold{X}_i)}\bigg)^{-1}\sum_{i=1}^{a^*}\dfrac{(1-A_i) I(Y_{(i)}\leq a)}{\widehat{\pi}(\bold{X}_i)}\\
    &=\widehat{F}^{(0)}(a),
\end{aligned}
\end{equation}
where the third step in (\ref{F1-monotone}) and (\ref{F0-monotone}) is due to that $I(Y_{(i)}\leq b ) = 0$ for $i = b^{*}+1,b^{*}+2,\dots,n$ and the fifth step in (\ref{F1-monotone}) and (\ref{F0-monotone}) is due to that $I(Y_{(i)}\leq a) = I(Y_{(i)}\leq b) = 1$ for $i = 1,2,\dots,a^*$. Then,  (\ref{F1-monotone}) and (\ref{F0-monotone}) imply that the CDFs are non-decreasing. Consequently, by (\ref{CDF-1-to1}) to (\ref{F0-monotone}), we verify that the proposed estimators (\ref{est-general treatment effect}) and (\ref{9}) are proper CDFs.
{\section{{Proof of Theorem \ref{recovery}}}}
We prove the variable selection consistency here. As Theorem \ref{recovery} is a statement with respect to both $\widehat{V}$ and $\widehat{E}$, we tackle the sparsity recoveries of $\widehat{V}$ and $\widehat{E}$ separately and eventually integrate the results to arrive at the desired theorem.\\
\underline{Part 1: \textit{Inference for }$\widehat{V}$}\\
We aim to show
\begin{equation}\label{sparsity-v}
    P(\widehat{V}= V)\to1\text{ as }n\to\infty.
\end{equation}
 Let the true coefficient be $\boldsymbol{\beta} = \bigg(\boldsymbol{\beta}_V^\top,\boldsymbol{\beta}_{V^c}^\top\bigg)^\top$ where $\boldsymbol{\beta}_V$ is the vector comprising coefficients of informative confounders with cardinality $|\boldsymbol{\beta}_V| = k$ and $\boldsymbol{\beta}_{V^c}$ contains coefficients of non-informative confounders and hence is $\bold{0}_{(p-k)}.$ Moreover, $\widehat{\boldsymbol{\beta}} = \bigg(\widehat{\boldsymbol{\beta}}_V^\top,\widehat{\boldsymbol{\beta}}_{V^c}^\top\bigg)^\top$ is the corresponding estimated version of $\boldsymbol{\beta}$. Let $\widehat{\bold{z}} =\bigg(\widehat{\bold{z}}_V^\top,\widehat{\bold{z}}_{V^C}^\top\bigg)^\top$ be the dual vector of $\widehat{\boldsymbol{\beta}}$, where the sub-vectors $\widehat{\bold{z}}_V^\top$ and $\widehat{\bold{z}}_{V^C}^\top$ are the dual vectors of $\widehat{\boldsymbol{\beta}}_V$ and $\widehat{\boldsymbol{\beta}}_{V^C}$, respectively, and the $j$th component is $\widehat{z}_j = \text{sign}({{\widehat{\beta}}}_j)$ if ${\widehat{\beta}}_j\not=0$ and $|z_j|\leq1$ otherwise. Also, any pair of $(\widehat{\boldsymbol{\beta}},\widehat{\bold{z}})$ satisfies the zero sub-gradient condition:
 \begin{equation}\label{zero-gradient}
     -\frac{1}{n}\bold{X}^\top(\bold{Y}-\bold{X}\widehat{\boldsymbol{\beta}})+ \lambda_1\widehat{\bold{z}} = \bold{0}.
 \end{equation}
According to Hastie et al. (2015), condition (A1) together with primal dual witness (PDW) procedure can be used to examine the sparsity of $\widehat{\boldsymbol{\beta}}$. The PDW procedure is stated as follows:
\begin{enumerate}
    \item Set $\widehat{\boldsymbol{\beta}}_{V^C} = 0$.
    \item Determine $\widehat{\bold{z}}_{V^C}$ according to (\ref{zero-gradient}).
    \item Verify whether or not
    \begin{equation}\label{z-vector-v}
        ||\widehat{\bold{z}}_{V^C}||_\infty <1.
    \end{equation}
\end{enumerate}
As discussed in Lemma 11.2. of Hastie et al. (2015, p.307), if (\ref{z-vector-v}) is true, then the vector $(\widehat{\boldsymbol{\beta}}_V,\bold{0})$ is the unique optimal solution to the lasso program, which guarantees (\ref{sparsity-v}). Therefore, the main focus of remaining proof is to show (\ref{z-vector-v}). Following the relationship in (\ref{zero-gradient}) with $\boldsymbol{\Theta}$ fixed, we have that
\begin{equation}\label{z}
    \begin{aligned}
    \widehat{\bold{z}} &= \frac{1}{n\lambda_1}\bold{X}(\bold{Y}-\bold{X}\widehat{\boldsymbol{\beta}})\\
    &=  \frac{1}{n\lambda_1} (\bold{X}^\top\bold{Y} - \bold{X}^\top\bold{X}\widehat{\boldsymbol{\beta}})\\
    &=\frac{1}{n\lambda_1} \bigg\{\bold{X}^\top(\bold{X}\boldsymbol{\beta}+ \bold{X}^{\circ2}\text{vec}(\boldsymbol{\Theta}) + \boldsymbol{\varepsilon}) - \bold{X}^\top\bold{X}\widehat{\boldsymbol{\beta}}\bigg\}\\
    &= \frac{1}{n\lambda_1} (\bold{X}^\top\bold{X}\boldsymbol{\beta}+ \bold{X}^\top \bold{X}^{\circ2}\text{vec}(\boldsymbol{\Theta}) +\bold{X}^\top\boldsymbol{\varepsilon}-\bold{X}^\top\bold{X}\widehat{\boldsymbol{\beta}}) \\
    &=\frac{1}{n\lambda_1} \bigg\{(\bold{X}^\top\bold{X})(\boldsymbol{\beta}-\widehat{\boldsymbol{\beta}})+\bold{X}^\top\bold{X}^{\circ2}\text{vec}(\boldsymbol{\Theta})+\bold{X}^\top\boldsymbol{\varepsilon}\bigg\}.
\end{aligned}
\end{equation}
Since $\widehat{\bold{z}}_{V^C}$ is the sub-vector of $\widehat{\bold{z}}$, we write (\ref{z}) in the block matrix form, so that $\widehat{\bold{z}}_{V}$ and $\widehat{\bold{z}}_{V^C}$ can be expressed as
$$
 \begin{bmatrix}
        \widehat{\bold{z}}_V\\
        \widehat{\bold{z}}_{V^C}
    \end{bmatrix} = \frac{1}{n\lambda_1}\Bigg\{\begin{bmatrix}
        \bold{X}_V^\top \bold{X}_{V} & \bold{X}_{V}^\top \bold{X}_{V^c}\\
        \bold{X}_{V^C}^\top \bold{X}_{V} & \bold{X}_{V^C}^\top \bold{X}_{V^C}
    \end{bmatrix} \begin{bmatrix}
            \boldsymbol{\beta}_V - \widehat{\boldsymbol{\beta}}_V \\
            \bold{0}_{(p-k)}
        \end{bmatrix} +\begin{bmatrix}
            \bold{X}_{V}^\top\bold{X}^{\circ2}\text{vec}(\boldsymbol{\Theta}) \\
            \bold{X}_{V^C}^\top\bold{X}^{\circ2}\text{vec}(\boldsymbol{\Theta})
        \end{bmatrix}+\begin{bmatrix}
            \bold{X}_V^\top\boldsymbol{\varepsilon} \\
            \bold{X}_{V^C}^\top\boldsymbol{\varepsilon}
        \end{bmatrix}\Bigg\}.
$$
It also implies that
\begin{equation}\label{dual-vector-v}
    \widehat{\bold{z}}_{V} = \frac{1}{n\lambda_1}\bigg(\bold{X}_{V}^\top \bold{X}_{V}( \boldsymbol{\beta}_V - \widehat{\boldsymbol{\beta}}_V)+\bold{X}_{V}^\top\bold{X}^{\circ2}\text{vec}(\boldsymbol{\Theta})+\bold{X}_{V}^\top\boldsymbol{\varepsilon}\bigg)
\end{equation}
and
\begin{equation}\label{dual-vector-vc}
    \widehat{\bold{z}}_{V^C} =\frac{1}{n\lambda_1}\bigg( \bold{X}_{V^C}^\top \bold{X}_{V}( \boldsymbol{\beta}_V - \widehat{\boldsymbol{\beta}}_V)+\bold{X}_{V^C}^\top\bold{X}^{\circ2}\text{vec}(\boldsymbol{\Theta})+\bold{X}_{V^C}^\top\boldsymbol{\varepsilon}\bigg).
\end{equation}
By (\ref{dual-vector-v}) and {Condition (A1)}, the difference $\boldsymbol{\beta}_V - \widehat{\boldsymbol{\beta}}_V$ can  be expressed as 
\begin{equation}\label{beta diff}
    \boldsymbol{\beta}_V - \widehat{\boldsymbol{\beta}}_V = n\lambda_1 (\bold{X}_{V}^\top \bold{X}_{V})^{-1}\text{sign}(\boldsymbol{\beta}_V) - (\bold{X}_{V}^\top \bold{X}_{V})^{-1}\bold{X}_{V}\bold{X}^{\circ2}\text{vec}(\boldsymbol{\Theta})- (\bold{X}_{V}^\top \bold{X}_{V})^{-1}\bold{X}_{V}^\top\boldsymbol{\varepsilon}.
\end{equation}
Replacing the $( \boldsymbol{\beta}_V - \widehat{\boldsymbol{\beta}}_V)$ in (\ref{dual-vector-vc}) by (\ref{beta diff}) yields that
$$
\begin{aligned}
    \widehat{\bold{z}}_{V^C} = \bold{X}_{V^C}^\top \bold{X}_{V}(\bold{X}_{V}^\top \bold{X}_{V})^{-1}\text{sign}(\boldsymbol{\beta}_V) &-\frac{\bold{X}_{V^C}^\top \bold{X}_{V}(\bold{X}_{V}^\top \bold{X}_{V})^{-1}\bold{X}_{V}\bold{X}^{\circ2}\text{vec}(\boldsymbol{\Theta})}{n\lambda_1}+\frac{\bold{X}_{V^C}^\top\bold{X}^{\circ2}\text{vec}(\boldsymbol{\Theta})}{n\lambda_1} \\
    & -\frac{\bold{X}_{V^C}^\top \bold{X}_{V}(\bold{X}_{V}^\top \bold{X}_{V})^{-1}\bold{X}_{V}^\top\boldsymbol{\varepsilon}}{n\lambda_1}+\frac{\bold{X}_{V^C}^\top\boldsymbol{\varepsilon}}{n\lambda_1}\\
    = {\bold{X}_{V^C}^\top \bold{X}_{V}(\bold{X}_{V}^\top \bold{X}_{V})^{-1}\text{sign}(\boldsymbol{\beta}_V)}&+\bold{X}_{V^C}^\top[\bold{I}- \bold{X}_{V}(\bold{X}_{V}^\top \bold{X}_{V})^{-1}\bold{X}_{V}]\bigg(\frac{\bold{X}^{\circ2}\text{vec}(\boldsymbol{\Theta})}{n\lambda_1}\bigg)\\
    &+\bold{X}^\top_{V^C}[\bold{I} - \bold{X}_V(\bold{X}_V^\top\bold{X}_V)^{-1}\bold{X}_V^\top]\bigg(\frac{\boldsymbol{\varepsilon}}{n\lambda_1}\bigg) \\
    ={\bold{X}_{V^C}^\top \bold{X}_{V}(\bold{X}_{V}^\top \bold{X}_{V})^{-1}\text{sign}(\boldsymbol{\beta}_V)} &+ {\bold{X}_{V^C}^\top[\bold{I}- \bold{X}_{V}(\bold{X}_{V}^\top \bold{X}_{V})^{-1}\bold{X}_{V}] \bigg(\frac{\bold{X}^{\circ2}\text{vec}(\boldsymbol{\Theta})+\boldsymbol{\varepsilon}}{n\lambda_1}\bigg)}\\
    \triangleq \boldsymbol{\mu} + \boldsymbol{\Psi}.\qquad\qquad\qquad\qquad\;\;\;\;&
\end{aligned}
$$
By triangle inequality,
\begin{equation}\label{triangle - zvc}
    \|\widehat{\bold{z}}_{V^C}\|_{\infty} \leq \|\boldsymbol{\mu}\|_{\infty} + \|\boldsymbol{\Psi}\|_{\infty}. 
\end{equation}
With regularity condition (A2), $\|\boldsymbol{\mu}\|_{\infty}\leq 1-\gamma_1$. Hence, we only have to show that $\|\boldsymbol{\Psi}\|_{\infty}\leq\gamma_1$ with high probability. Let $ \Pi_{V^\perp}(\bold{X}) \triangleq \bold{I}- \bold{X}_{V}(\bold{X}_{V}^\top \bold{X}_{V})^{-1}\bold{X}_{V}$ be an $n\times n$ orthogonal projection matrix, then one can show that for $r\in V^C$,
$$
\boldsymbol{\Psi}_r = \bold{X}^\top_r \Pi_{V^\perp}(\bold{X})\bigg(\frac{\bold{X}^{\circ2}\vecTheta+ \boldsymbol{\varepsilon}}{n\lambda_1}\bigg),
$$
which is the $r$th component in $\boldsymbol{\Psi}$, is a Gaussian random variable with mean $\xi\triangleq\bold{X}^{\circ2}\vecTheta$ and variance at most $\dfrac{\sigma^2 K_{\text{clm}}^2}{n\lambda_1^2}$. Without loss of generality, let $\boldsymbol{\Psi}_{\text{max}}$ be a normal random variable with mean $\bold{X}^{\circ2}\vecTheta$ and possibly maximal variance $\dfrac{\sigma^2 K_{\text{clm}}^2}{n\lambda_1^2}$. {That is, for any $r\in\{1,\dots,p\}$, $\text{Var}(\boldsymbol{\Psi}_{\text{max}})\ge\text{Var}(\boldsymbol{\Psi}_r)$.} Applying Lemma \ref{tail bound} gives that
\begin{equation}\label{tail bound of Psi-max}
    P\bigg(|\boldsymbol{\Psi}_{\text{max}}|\ge \gamma_1\bigg)\leq 2\text{exp}\bigg(-\dfrac{(\gamma_1-\xi)^2 n \lambda_1^2}{2\sigma^2 K_{\text{clm}}^2}\bigg).
\end{equation}
The upper bound in (\ref{tail bound of Psi-max}) approaches to $0$ as $n\to\infty$, which implies 
$$
    P\bigg(|\boldsymbol{\Psi}_{\text{max}}|\leq \gamma_1\bigg) \ge 1 \text{ as }n\to\infty.
$$
Introducing the union bound further yields that
\begin{equation}\label{Psi-inf-norm}
    P\bigg(\|\boldsymbol{\Psi}\|_{\infty} \ge{\gamma_1}\bigg) \leq 
    2(p-k)\text{exp}\bigg(-\dfrac{(\gamma_1-\xi)^2 n \lambda_1^2}{2\sigma^2 K_{\text{clm}}^2}\bigg),
\end{equation}
ensuring that $\|\boldsymbol{\Psi}\|_{\infty}\leq\gamma_1$ as $n\to\infty$. Combining $\|\boldsymbol{\mu}\|_{\infty}\leq 1-\gamma_1$ from regularity condition (A2) and $\|\boldsymbol{\Psi}\|_{\infty}$ with (\ref{triangle - zvc}) gives (\ref{z-vector-v}), ensuring that the sparsity recovery is guaranteed for $\widehat{V}$.
\\
\underline{Part 2: \textit{Inference for }$\widehat{E}$}\\
In this part, our goal is to show 
$$
    P(\widehat{E}= E)\to1\text{ as }n\to\infty.
$$
The procedure in Part 1 is repeated with the target of inference changed to $\bold{\Theta}$. Let the vector of true coefficients be $\vecTheta = \bigg( \text{vec}(\boldsymbol{\Theta}_E)^\top, \text{vec}(\boldsymbol{\Theta}_{E^C})^\top \bigg)^\top$ where $\text{vec}(\boldsymbol{\Theta}_E)$ is a $m$-dimensional nonzero vector that contains the coefficients of informative interaction terms and $\text{vec}(\boldsymbol{\Theta}_{E^C})$ consists of those of non-informative interaction terms and is equal to $\bold{0}_{{p\choose2} - m}$.{ Moreover, $\text{vec}(\boldsymbol{\widehat{\Theta}}) = \bigg(\text{vec}(\boldsymbol{\widehat{\Theta}}_E)^\top, \text{vec}(\boldsymbol{\widehat{\Theta}}_{E^C})^\top\bigg)^\top $ is the corresponding estimated version of $\text{vec}(\boldsymbol{\Theta})$. Let $\widehat{\bold{z}}^\prime = \bigg(\widehat{\bold{z}}^\prime_{E},\widehat{\bold{z}}^\prime_{E^C}\bigg)$ be the dual vector of $\text{vec}(\widehat{\boldsymbol{\Theta}})$, where the sub-vectors $\widehat{\bold{z}}^\prime_{E}$ and $  \widehat{\bold{z}}^\prime_{E^C}$ are the dual vectors of $\text{vec}(\boldsymbol{\widehat{\Theta}}_E)$ and $\text{vec}(\boldsymbol{\widehat{\Theta}}_{E^C})$, respectively, and the $j$th component is $\widehat{z}^\prime_j$ = \text{sign}$(\widehat{\theta}_j)$ if $\widehat{\theta}_j\not=0$ and $|\widehat{z}^\prime_j|\leq1$ otherwise.} The zero sub-gradient condition with respect to $\vecTheta$ is
$$
-\frac{1}{n}(\bold{X}^{\circ2})^\top(\bold{Y}-\bold{X}^{\circ2}\vecThetah)+\lambda_2\widehat{\bold{z}}^\prime = \bold{0}.
$$
Similar to the procedure in Part 1, the condition
\begin{equation}\label{z ec}
\|\widehat{\bold{z}}^\prime_{E^C}\|_{\infty} \leq1
\end{equation}
is sufficient to attain the sparsity for $\widehat{E}$ and therefore is of primary attention in this part. With the similar procedure shown in Part 1, we express the dual vector as 
$$
    \widehat{\bold{z}}^\prime_{E^C} ={\bold{X}^{\circ2\top}_{E^C}}\bold{X}^{\circ2}_E(\bold{X}^{\circ2\top}_E\bold{X}^{\circ2}_E)^{-1}\text{sign}(\text{vec}(\boldsymbol{\Theta}_E)) + \bold{X}^{\circ2\top}_{E^C}[\bold{I}- \bold{X}^{\circ2}_E(\bold{X}^{\circ2\top}_E \bold{X}^{\circ2}_E)^{-1}\bold{X}^{\circ2}_E]\bigg(\frac{\bold{X}\boldsymbol{\beta}+\boldsymbol{\epsilon}}{n\lambda_2}\bigg).
$$
In a similar manner with Part 1, we can apply the triangle inequality to this vector. Then, by utilizing Condition (A2) and Lemma \ref{tail bound}, one can show that (\ref{z ec}) holds with high probability with $n\to\infty$. Therefore, the sparsity consistency for $\widehat{E}$ is assured as $n\to\infty$. 
\\
\underline{Part 3: \textit{Inference for }$\widehat{G}$}\\
As $G = (V,E)$, the statement $\widehat{G}= G$ can be viewed as $\widehat{V}= V$ and $\widehat{E}= E$. On one hand, By Bonferroni's inequality, we have
\begin{equation}\label{bonferroni}
    P(\widehat{G}= G) = P(\widehat{V}= V, \widehat{E}= E) \ge P(\widehat{V}= V)+P(\widehat{E}= E) -1.
\end{equation}
On the other hand, the inclusion relationship between events implies that
\begin{equation}\label{inclusion}
    P(\widehat{G}= G) \leq \min{\bigg\{P(\widehat{V}= V), P(\widehat{E}= E)\bigg\}}.
\end{equation}
Combining (\ref{bonferroni}) and (\ref{inclusion}) yields that (e.g. Chen and Yi, 2021)
\begin{equation}
    P(\widehat{V}= V)+P(\widehat{E}= E) -1 \leq P(\widehat{G}= G) \leq \min{\bigg\{P(\widehat{V}= V), P(\widehat{E}= E)\bigg\}}.
\end{equation}
By the squeeze theorem, we have $P(\widehat{G}= G)$ as $n\to\infty$. Consequently, the sparsity for $G$ is verified.$\hfill\square$
\section{{Proof of Theorem \ref{asymptotic for our estimator}}}\label{appendix:e}

In this section, we consider the limiting distribution where the underlying network structure $G$ is correctly specified by previous variable selection procedure. A new design matrix of size $n\times (1+k+m)$ is defined as $\bold{X}^* = \bigg[\bold{1}\;\bold{X}_{\widehat{V}} \;\bold{X}_{\widehat{E}}\bigg]$ and the submatrices $\bold{X}_{\widehat{V}}$ and $\bold{X}_{\widehat{E}}$ consist of the chosen main and interaction effects, respectively. To derive the asymptotic results for our estimator, we first expand our estimator in terms of Hadamard derivative to determine the dominant term. Secondly, as the dominant term consists of a sequence of estimation in different parameters, we adopt M-estimation (Stefanski and
Boos, 2002) to investigate the asymptotic result of the dominant term.\\
\underline{Part 1: \textit{Expansion and Determining the Dominant Term}}\\
Let $U^{(a)} = F^{(a)}\circ (F^{(a)})^{-1} $, which follows the uniform distribution function as defined in (\ref{uniform distribution}), and $\widehat{U}^{(a)}= \widehat{F}^{(a)}\circ (F^{(a)})^{-1}$ for $a=0,1$. By the similar derivation of (\ref{expansion}) in Lemma \ref{asymptotic normallity stat fun}, we have that
\begin{equation}\label{F1-expansion}
    \begin{aligned}
        &\;\;\;\;\sqrt{n}\bigg\{T(\widehat{F}^{(1)})-T({F}^{(1)})\bigg\} \\
        &= \sqrt{n}\rho_U^\prime(\widehat{U}^{(1)}-U^{(1)}) +\sqrt{n}\text{Rem}(\widehat{U}^{(1)}-U^{(1)})\\
        &=\sqrt{n}\Bigg[\rho_U^\prime\bigg((\widehat{F}^{(1)}-{F^{(1)}})\circ \{F^{(1)}\}^{-1}\bigg)\Bigg]+\sqrt{n}\textup{Rem}(\widehat{U}^{(1)}-U^{(1)})\\
       &=\sqrt{n}\Bigg[\rho_U^\prime\bigg(\bigg(\sum_{i=1}^n\frac{A_i}{\widehat{\pi}(\bold{X}^*_i)}\bigg)^{-1}\sum_{i=1}^n\bigg(\frac{A_i\delta_{Y_i}}{\widehat{\pi}(\bold{X}^*_i)}-{F}^{(1)}\bigg)\circ \{F^{(1)}\}^{-1}\bigg)\Bigg]+\sqrt{n}\textup{Rem}(\widehat{U}^{(1)}-U^{(1)})\\
       &=\sqrt{n}\bigg(\sum_{i=1}^n\frac{A_i}{\widehat{\pi}(\bold{X}^*_i)}\bigg)^{-1}\bigg(\sum_{i=1}^n\frac{A_i\phi_{F^{(1)}}(Y_i)}{\widehat{\pi}(\bold{X}^*_i)}\bigg)+\sqrt{n}\textup{Rem}(\widehat{U}^{(1)}-U^{(1)}).
    \end{aligned}
\end{equation}
Similar derivation for $\sqrt{n}\bigg\{T(\widehat{F}^{(0)})-T({F}^{(0)})\bigg\}$ gives that
\begin{equation}\label{F0-expansion}
  \begin{aligned}
        \sqrt{n}\bigg\{T(\widehat{F}^{(0)})-T({F}^{(0)})\bigg\} =&\sqrt{n}\bigg(\sum_{i=1}^n\frac{1-A_i}{1-\widehat{\pi}(\bold{X}^*_i)}\bigg)^{-1}\bigg(\sum_{i=1}^n\frac{(1-A_i)\phi_{F^{(0)}}(Y_i)}{1-\widehat{\pi}(\bold{X}^*_i)}\bigg)\\
        &+\sqrt{n}\textup{Rem}(\widehat{U}^{(0)}-U^{(0)}).
  \end{aligned}
\end{equation}
Combining (\ref{F1-expansion}) and (\ref{F0-expansion}) leads to that
\begin{equation}\label{F1F0-expansion}
\begin{aligned}
        &\;\;\;\;\sqrt{n}\Bigg[\bigg(T(\widehat{F}^{(1)})-T(\widehat{F}^{(0)})\bigg)-\bigg(T({F}^{(1)})-T({F}^{(0)})\bigg)\Bigg] \\
        &=  \sqrt{n}\bigg(\sum_{i=1}^n\frac{A_i}{\widehat{\pi}(\bold{X}^*_i)}\bigg)^{-1}\bigg(\sum_{i=1}^n\frac{A_i\phi_{F^{(1)}}(Y_i)}{\widehat{\pi}(\bold{X}^*_i)}\bigg)\\
    &\;\;\;\;-\sqrt{n}\bigg(\sum_{i=1}^n\frac{1-A_i}{1-\widehat{\pi}(\bold{X}^*_i)}\bigg)^{-1}\bigg(\sum_{i=1}^n\frac{(1-A_i)\phi_{F^{(0)}}(Y_i)}{1-\widehat{\pi}(\bold{X}^*_i)}\bigg)\\
    &\;\;\;\;+\sqrt{n}\bigg(\textup{Rem}(\widehat{U}^{(1)}-U^{(1)})-\textup{Rem}(\widehat{U}^{(0)}-U^{(0)})\bigg).
\end{aligned}
\end{equation}
Since $\text{Rem}(\widehat{U}^{(a)}-U^{(a)})$ in (\ref{F1F0-expansion}) converges to $0$ in probability as $n\to\infty$ for $a=0,1$ (Fernholz, 1983), it suffices to analyze
\begin{equation}\label{dominant term}
    \sqrt{n}\Bigg[ \bigg(\sum_{i=1}^n\frac{A_i}{\widehat{\pi}(\bold{X}_i)}\bigg)^{-1}\sum_{i=1}^n \frac{A_i\phi_{F^{(1)}}(Y_i)}{\widehat{\pi}(\bold{X}_i)} -\bigg(\sum_{i=1}^n\frac{1-A_i}{1-\widehat{\pi}(\bold{X}_i)}\bigg)^{-1}\sum_{i=1}^n \frac{(1-A_i)\phi_{F^{(0)}}(Y_i)}{1-\widehat{\pi}(\bold{X}_i)}\Bigg].
\end{equation}
We employ M-estimation to further investigate the asymptotic variance of (\ref{dominant term}).
\\
\underline{Part 2: \textit{Estimating Equations Construction}}

First, (\ref{logistic-estimator}) indicates that $\widehat{\boldsymbol{\eta}}$ must satisfy the following equation
\begin{equation}\label{logistic-regression-star-ee}
    \sum_{i=1}^n \psi_{\boldsymbol{\eta}}(A_i,\bold{X}^*_i\mid\widehat{\boldsymbol{\eta}}) \triangleq \sum_{i=1}^n \bigg[A_i-\pi(\bold{X}^*_i\mid\widehat{\boldsymbol{\eta}})\bigg] \bold{X}^{*\top}_i= \bold{0}_{1+k+m}.
\end{equation}
Next, we consider the estimating equation for IPW-CDF under $a\in\{0,1\}$, where are given by
\begin{equation}
    \psi_{1}(Y_i,\bold{X}^*_i,A_i\mid \boldsymbol{\eta},\kappa_1) \triangleq \dfrac{A_i\bigg(\phi_{F^{(1)}}(Y_i) -\kappa_1\bigg)}{\pi(\bold{X}^*_i\mid\boldsymbol{\eta})}
\end{equation}
and
\begin{equation}\label{E.4}
    \psi_{0}(Y_i,\bold{X}^*_i,A_i\mid \boldsymbol{\eta},\kappa_0) \triangleq \frac{(1-A_i)\bigg(\phi_{F^{(0)}}(Y_i) -\kappa_0\bigg)}{1-\pi(\bold{X}^*_i\mid\boldsymbol{\eta})},
\end{equation}
where $\kappa_1$ and $\kappa_0$ are two parameters and are used to ensure that $E[\psi_{1}(Y,\bold{X}^*,A\mid \boldsymbol{\eta},\kappa_1)] = 0$ and $E[\psi_{0}(Y,\bold{X}^*,A\mid \boldsymbol{\eta},\kappa_0)] = 0$. Moreover, the solution to $\sum\limits_{i=1}^n\psi_{1}(Y_i,\bold{X}_i^*,A_i\mid\boldsymbol{\eta},\kappa_1) =0$ and  $\sum\limits_{i=1}^n\psi_{0}(Y_i,\bold{X}_i^*,A_i\mid\boldsymbol{\eta},\kappa_0) =0$ are denoted as $\widehat{\kappa}_1$ and  $\widehat{\kappa}_0$, respectively. One may check that 
$$
\widehat{\kappa}_1 = \bigg(\sum_{i=1}^n\frac{A_i}{\widehat{\pi}(\bold{X}^*_i)}\bigg)^{-1}\bigg(\sum_{i=1}^n\frac{A_i\phi_{F^{(1)}}(Y_i)}{\widehat{\pi}(\bold{X}^*_i)}\bigg)
$$
and 
$$
\widehat{\kappa}_0 = \bigg(\sum_{i=1}^n\frac{1-A_i}{1-\widehat{\pi}(\bold{X}^*_i)}\bigg)^{-1}\bigg(\sum_{i=1}^n\frac{(1-A_i)\phi_{F^{(0)}}(Y_i)}{1-\widehat{\pi}(\bold{X}^*_i)}\bigg).
$$
\\
\underline{Part 3: \textit{M-Estimation}}\\
Following the procedure in Stefanski and Boos (2002) with the estimating equations (\ref{logistic-regression-star-ee})- (\ref{E.4}) in Part 2, we define $\boldsymbol{\vartheta} \triangleq (\boldsymbol{\eta}^\top,\kappa_1,\kappa_0)^\top$ and 
\begin{equation}\label{EE vartheta}
    \boldsymbol{\psi}(Y_i,\bold{X}^*_i,A_i\mid\boldsymbol{\vartheta}) \triangleq \begin{bmatrix}
    \psi_{{\boldsymbol{\eta}}}(A_i,\bold{X}^*_i\mid\boldsymbol{\eta}) \\
    \psi_{1}(Y_i,\bold{X}^*_i,A_i\mid \boldsymbol{\eta},\kappa_1) \\
    \psi_{0}(Y_i,\bold{X}^*_i,A_i\mid \boldsymbol{\eta},\kappa_0)
\end{bmatrix} = \begin{bmatrix}
    \bigg[A_i-\pi(\bold{X}^*_i\mid{\boldsymbol{\eta}})\bigg] \bold{X}^{*\top}_i \\
    \dfrac{A_i\bigg(\phi_{F^{(1)}}(Y_i) -\kappa_1\bigg)}{\pi(\bold{X}^*_i\mid\boldsymbol{\eta})} \\
    \dfrac{(1-A_i)\bigg(\phi_{F^{(0)}}(Y_i) -\kappa_0\bigg)}{1-\pi(\bold{X}^*_i\mid\boldsymbol{\eta})}
\end{bmatrix}.
\end{equation}
Since $\widehat{\boldsymbol{\vartheta}} \triangleq (\widehat{\boldsymbol{\eta}}^\top,\widehat{\kappa}_1,\widehat{\kappa}_0)^\top$ is the solution of  $\sum_{i=1}^n\boldsymbol{\psi}(Y_i,\bold{X}^*_i,A_i\mid\boldsymbol{\vartheta})=\bold{0}_{3+k+m}$, then under suitable conditions (Stefanski and Boos, 2002), we have that
\begin{equation}\label{M-estimation asymptotic normality}
\sqrt{n}(\widehat{\boldsymbol{\vartheta}}-\boldsymbol{\vartheta}) \overset{d}{\longrightarrow} \text{N}(0,\bold{V}(\boldsymbol{\vartheta})) \text{ as } n\to\infty,
\end{equation}
where $\bold{V}(\boldsymbol{\vartheta}) = \bold{A}^{-1}(\boldsymbol{\vartheta})\bold{B}(\boldsymbol{\vartheta})\{\bold{A}^{-1}(\boldsymbol{\vartheta})\}^\top$ with the matrices
\begin{equation}\label{B matrix}
    \begin{aligned}
         \bold{B}(\boldsymbol{\vartheta}) &= E\Bigg[\boldsymbol{\psi}(Y_i,\bold{X}^*_i,A_i\mid\boldsymbol{\vartheta})\bigg\{\boldsymbol{\psi}(Y_i,\bold{X}^*_i,A_i\mid\boldsymbol{\vartheta})\bigg\}^\top\Bigg]\\
        &= E\left[\begin{array}{c|cc}
    \psi_{\boldsymbol{\eta}}\psi_{\boldsymbol{\eta}}^\top  &\psi_{\boldsymbol{\eta}}\psi_1 &\psi_{\boldsymbol{\eta}}\psi_0\\[0.25cm]
            \hline
            \psi_1\psi_{\boldsymbol{\eta}}^\top &\psi_1\psi_1 &\psi_1\psi_0\\[0.25cm]
            \psi_0\psi_{\boldsymbol{\eta}}^\top &\psi_0\psi_1 &\psi_0\psi_0
        \end{array}\right]\\
    &\triangleq \begin{bmatrix}
        {\bf b}_{11} &\bold{b}_{21}^\top\\
        \bold{b}_{21} &\bold{b}_{22}
    \end{bmatrix}
    \end{aligned}
\end{equation}
and
\begin{equation}\label{A matrix}
    \begin{aligned}
     \bold{A}(\boldsymbol{\vartheta}) &= - E\bigg[\dfrac{\partial}{\partial\boldsymbol{\vartheta}}\boldsymbol{\psi}(Y,\bold{X}^*,A\mid\boldsymbol{\vartheta})\bigg] \\
    &=-E \begin{bmatrix}
    \dfrac{\partial}{\partial\boldsymbol{\eta}}\psi_{\boldsymbol{\eta}}(Y,\bold{X}^*,A\mid \boldsymbol{\eta}) & \bold{0}_{1+k+m}&\bold{0}_{1+k+m}\\
     \dfrac{\partial}{\partial\boldsymbol{\eta}}\psi_{1}(Y,\bold{X}^*,A\mid \boldsymbol{\eta},\kappa_1) & \dfrac{\partial}{\partial\kappa_1}\psi_{1}(Y,\bold{X}^*,A\mid \boldsymbol{\eta},\kappa_1)&0\\
      \dfrac{\partial}{\partial\boldsymbol{\eta}} \psi_{0}(Y,\bold{X}^*,A\mid \boldsymbol{\eta},\kappa_0)&0 &\dfrac{\partial}{\partial\kappa_0}\psi_0(Y,\bold{X}^*,A\mid \boldsymbol{\eta},\kappa_0)
\end{bmatrix}\\
&= -E\left[\begin{array}{c|cc}
   \pi(\bold{X}^*\mid\boldsymbol{\eta})\bigg(1-\pi(\bold{X}^*\mid\boldsymbol{\eta})\bigg) \bold{X}^{*\top}\bold{X}^{*}& \bold{0}_{1+k+m}&\bold{0}_{1+k+m}\\[0.25cm]
   \hline
     \dfrac{1-\pi(\bold{X}^*\mid\boldsymbol{\eta})}{\pi(\bold{X}^*\mid\boldsymbol{\eta})}A\bigg(\phi_{F^{(1)}}(Y_i) -\kappa_1\bigg)\bold{X}^*& -1&0\\[0.25cm]
    -\dfrac{\pi(\bold{X}^*\mid\boldsymbol{\eta})}{1-\pi(\bold{X}^*\mid\boldsymbol{\eta})}(1-A)\bigg(\phi_{F^{(0)}}(Y_i) -\kappa_0\bigg)\bold{X}^*&0 & -1
\end{array}\right]\\
&\triangleq \begin{bmatrix}
    {\bf a}_{11} & \bold{0}_{(1+k+m)\times2}\\
    {\bf a}_{21} & {\bf I}_2
\end{bmatrix}
\end{aligned}
\end{equation} 
with the inverse matrix
\begin{equation}
\begin{aligned}
          \bold{A}^{-1}(\boldsymbol{\vartheta}) = \begin{bmatrix}
    {\bf a}_{11}^{-1} &\bold{0}_{(1+k+m)\times2}  \\
     -{\bf a}_{11}^{-1}{\bf a}_{21} &\bold{I}_2 
\end{bmatrix}
\end{aligned}
\end{equation}
and $\bold{I}_2$ being a $2\times2$ identity matrix. It suggests that $\bold{V}(\boldsymbol{\vartheta})$ can be expressed as
$$
\begin{aligned}
    \bold{V}(\boldsymbol{\vartheta}) &= \bold{A}^{-1}(\boldsymbol{\vartheta})  \bold{B}(\boldsymbol{\vartheta})\{ \bold{A}^{-1}(\boldsymbol{\vartheta})\}^\top\\
    &=\begin{bmatrix}
    {\bf a}_{11}^{-1} &\bold{0}_{(1+k+m)\times2}  \\
     -{\bf a}_{11}^{-1}{\bf a}_{21} &{\bf I}_2 
\end{bmatrix}\begin{bmatrix}
    {\bf a}_{11} & {\bf b}_{21}^\top\\
    {\bf b}_{21} & {\bf b}_{22}
\end{bmatrix} \begin{bmatrix}
    {\bf a}_{11}^{-1} &  -{\bf a}_{21}^\top {\bf a}_{11}^{-1} \\
    \bold{0}_{2\times(1+k+m)} &{\bf I}_2 
\end{bmatrix}\\
&=\begin{bmatrix}
    {\bf a}_{11}^{-1} & -{\bf a}_{11}^{-1} {\bf a}_{21}^\top  -{\bf a}_{11}^{-1}{\bf b}_{21}\\
    (-{\bf a}_{21}+{\bf b}_{21}){\bf a}_{11}^{-1} & ({\bf a}_{21}-{\bf b}_{21}){\bf a}_{11}^{-1}{\bf a}_{21}^\top-{\bf a}_{21}{\bf a}_{11}^{-1}{\bf b}_{21}^\top +{\bf b}_{22}
\end{bmatrix}.
\end{aligned}
$$

Now we thoroughly investigate the components of the matrices (\ref{B matrix}) and (\ref{A matrix}). Firstly, by Lemma 7.3.11 of Casella and Berger (2002), we have that ${\bf a}_{11} = {\bf b}_{11}= E\Bigg\{\pi(\bold{X}^*\mid\boldsymbol{\eta})\bigg(1-\pi(\bold{X}^*\mid\boldsymbol{\eta})\bigg) \bold{X}^{*\top}\bold{X}^{*}\Bigg\}$. Secondly,
$$
\begin{aligned}
    {\bf a}_{21} &\triangleq \Bigg[-E\bigg(\dfrac{\partial}{\partial\boldsymbol{\eta}} \psi_1\bigg) -E\bigg(\dfrac{\partial}{\partial\boldsymbol{\eta}} \psi_0\bigg)\Bigg]^{\top} \\
        &=\Bigg[E\bigg\{\frac{1-\pi(\bold{X}^*\mid\boldsymbol{\eta})}{\pi(\bold{X}^*\mid\boldsymbol{\eta})}A(\phi_{F^{(1)}}(Y)-\kappa_1)\bold{X}^*\bigg\} \;\;-E\bigg\{\frac{\pi(\bold{X}^*\mid\boldsymbol{\eta})}{1-\pi(\bold{X}^*\mid\boldsymbol{\eta})}(1-A)(\phi_{F^{(0)}}(Y)-\kappa_0)\bold{X}^*\bigg\}\Bigg]^\top\\
        &= \Bigg[E\bigg\{\bigg(1-\pi(\bold{X}^*\mid\boldsymbol{\eta})\bigg)\bold{X}^*(\phi_{F^{(1)}}(Y^{(1)})-\kappa_1)\bigg\}\;\; -E\bigg\{\pi(\bold{X}^*\mid\boldsymbol{\eta})\bold{X}^*(\phi_{F^{(0)}}(Y^{(0)})-\kappa_0)\bigg\}\Bigg]^\top.
\end{aligned}
$$

For the component $
    {\bf b}_{21} \triangleq \Bigg[E(\psi_1\psi_{\boldsymbol{\eta}}^\top)\;E(\psi_0\psi_{\boldsymbol{\eta}}^\top)\Bigg]^\top$in $\bold{B}(\boldsymbol{\vartheta})$, we observe that
\begin{equation}\label{psi1psieta}
    \begin{aligned}
    E(\psi_1\psi_{\boldsymbol{\eta}}^\top) &= E\Bigg[  \dfrac{A(\phi_{F^{(1)}}(Y)-\kappa_1)}{{\pi}(\bold{X}^*\mid \boldsymbol{\eta})}\bigg\{A-\pi(\bold{X}^*\mid\boldsymbol{\eta})\bold{X}^*\bigg\} \Bigg]\\
    &=E\Bigg\{  \dfrac{A(\phi_{F^{(1)}}(Y)-\kappa_1)\bold{X}^*}{{\pi}(\bold{X}^*\mid \boldsymbol{\eta})}-A(\phi_{F^{(1)}}(Y)-\kappa_1)\bold{X}^* \Bigg\}\\
    &=E\Bigg\{A(\phi_{F^{(1)}}(Y)-\kappa_1)\bold{X}^*\bigg(\frac{1}{\pi(\bold{X}^*\mid\boldsymbol{\eta})}-1\bigg)\Bigg\}\\
    &= E\Bigg\{A(\phi_{F^{(1)}}(Y)-\kappa_1)\bold{X}^*\bigg(\frac{1-\pi(\bold{X}^*\mid\boldsymbol{\eta})}{\pi(\bold{X}^*\mid\boldsymbol{\eta})}\bigg)\Bigg\}\\
    &=E\Bigg\{(\phi_{F^{(1)}}(Y^{(1)})-\kappa_1)\bold{X}^*\bigg({1-\pi(\bold{X}^*\mid\boldsymbol{\eta})}\bigg)\Bigg\}\\
    &= - E\bigg(\frac{\partial}{\partial\boldsymbol{\eta}}\psi_1\bigg)
\end{aligned}
\end{equation}
and 
$$
    \begin{aligned}
    E(\psi_0\psi_{\boldsymbol{\eta}}^\top)  = -E\bigg(\frac{\partial}{\partial\boldsymbol{\eta}}\psi_0\bigg) 
\end{aligned}
$$
derived by the similar step in (\ref{psi1psieta}). It suggests that ${\bf b}_{21} = {\bf a}_{21}$. Moreover, 
$$
\begin{aligned}
    {\bf b}_{22} &= E\begin{bmatrix}
        \psi_1\psi_1 & \psi_1\psi_0\\
        \psi_0\psi_1 & \psi_0\psi_0\\
    \end{bmatrix}\\
    &=\begin{bmatrix}
       E\bigg(\dfrac{A(\phi_{F^{(1)}}(Y)-\kappa_1)^2}{\pi(\bold{X}^*\mid\boldsymbol{\eta})^2}\bigg) & E\bigg\{\dfrac{A(1-A)(\phi_{F^{(1)}}(Y)-\kappa_1)(\phi_{F^{(0)}}(Y)-\kappa_0)}{\pi(\bold{X}^*\mid\boldsymbol{\eta})\{1-\pi(\bold{X}^*\mid\boldsymbol{\eta})\}}\bigg\}\\[0.5cm]
        E\bigg\{\dfrac{A(1-A)(\phi_{F^{(1)}}(Y)-\kappa_1)(\phi_{F^{(0)}}(Y)-\kappa_1)}{\pi(\bold{X}^*\mid\boldsymbol{\eta})\{1-\pi(\bold{X}^*\mid\boldsymbol{\eta})\}}\bigg\} &E\bigg[\dfrac{(1-A)(\phi(Y)-\kappa_0)^2}{\{1-\pi(\bold{X}^*\mid\boldsymbol{\eta})\}^2}\bigg]
    \end{bmatrix}\\
    &=\begin{bmatrix}
        E\bigg\{\dfrac{(\phi_{F^{(1)}}(Y^{(1)})-\kappa_1)^2}{\pi(\bold{X}^*\mid\boldsymbol{\eta})}\bigg\} &0 \\
        0 &E\bigg\{\dfrac{(\phi_{F^{(0)}}(Y^{(0)})-\kappa_0)^2}{1-\pi(\bold{X}^*\mid\boldsymbol{\eta})}\bigg\}
    \end{bmatrix},
\end{aligned}
$$
where the non-diagonal terms are $0$ since $A(1-A)=0$. 

By the property of the multivariate normal distribution, we immediately have that
$$
\sqrt{n}(\widehat{\boldsymbol{\eta}}-\boldsymbol{\eta}) \overset{d}{\longrightarrow}\text{N}(0,\mathcal{A}^{-1})
$$
as $n\to\infty$, where $\mathcal{A}\triangleq {\bf a}_{11}$. This gives the result (i). On the other hand, by (\ref{M-estimation asymptotic normality}), we have that, as $n\to\infty$,
\begin{equation}\label{kappas asymptotic normality}
    \sqrt{n}\Bigg\{\begin{pmatrix}
        \widehat{\kappa}_1\\
        \widehat{\kappa}_0
    \end{pmatrix}-\begin{pmatrix}
    \kappa_1\\
    \kappa_0
\end{pmatrix}\Bigg\} 
    \overset{d}{\longrightarrow} \text{N}({\bf 0}_2,{\bf v}_{22}),
\end{equation}
where
$$
\begin{aligned}
         {\bf v}_{22} & \triangleq ({\bf a}_{21}-{\bf b}_{21}){\bf a}_{11}^{-1}{\bf a}_{21}^\top-{\bf a}_{21}{\bf a}_{11}^{-1}{\bf b}_{21}^\top +{\bf b}_{22}\\
         &={\bf a}_{21}{\bf a}_{11}^{-1}{\bf a}_{21}^\top -{\bf b}_{21}{\bf a}_{11}^{-1}{\bf a}_{21}^\top -{\bf a}_{21}{\bf a}_{11}^{-1}{\bf b}_{12}^\top +{\bf b}_{22}\\
         &={\bf a}_{21}{\bf a}_{11}^{-1}{\bf a}_{21}^\top -{\bf b}_{21}{\bf a}_{11}^{-1}{\bf a}_{21}^\top -{\bf a}_{21}{\bf a}_{11}^{-1}{\bf b}_{21}^\top +{\bf b}_{21}{\bf a}_{11}^{-1}{\bf b}_{12}^\top-{\bf b}_{21}{\bf a}_{11}^{-1}{\bf b}_{21}^\top +{\bf b}_{22}\\
         &=({\bf a}_{21}-{\bf b}_{21}){\bf a}_{11}^{-1}({\bf a}_{21}-{\bf b}_{21})^\top-{\bf b}_{21}{\bf a}_{11}^{-1}{\bf b}_{21}^\top +{\bf b}_{22}\\
         &=-{\bf b}_{21}{\bf a}_{11}^{-1}{\bf b}_{21}^\top +{\bf b}_{22}.
    \end{aligned}
$$
Applying delta method to (\ref{kappas asymptotic normality}) with $\boldsymbol{\Delta}\triangleq (1,-1)^\top$, we have that, as $n\to\infty$,
\begin{equation}\label{delat-method-kappa}
    \sqrt{n}\bigg[(\widehat{\kappa}_1-\widehat{\kappa}_0) - (\kappa_1-\kappa_0)\bigg] \overset{d}{\longrightarrow}\text{N}\bigg(0,\mathcal{C}-\mathcal{B}\mathcal{A}^{-1}\mathcal{B}^\top\bigg),
\end{equation}
where $\mathcal{C}\triangleq \boldsymbol{\Delta}^\top {\bf b}_{22}\boldsymbol{\Delta}$ and $\mathcal{B}\triangleq\boldsymbol{\Delta}^\top{\bf b}_{21}$.

Recall that $\kappa_1$ and $\kappa_0$ are used to $E[\psi_{1}(Y,\bold{X}^*,A\mid \boldsymbol{\eta},\kappa_1)] = 0$ and $E[\psi_{0}(Y,\bold{X}^*,A\mid \boldsymbol{\eta},\kappa_0)] = 0$, respectively. Solving the first equation leads to
$$
\begin{aligned}
    \kappa_1 &= E\bigg\{\frac{A \phi_{F^{(1)}}(Y)}{\pi(\bold{X}^*)}\bigg\} \\
    &= E\Bigg\{\frac{1}{\pi(\bold{X}^*)}E\bigg( {A \phi_{F^{(1)}}(Y)} \mid \bold{X}^* \bigg)\Bigg\}\\
    &=E\Bigg\{\frac{E\bigg( {1\times \phi_{F^{(1)}}(Y^{(1)})} \mid \bold{X}^* ,A=1\bigg)P(A=1\mid\bold{X}^*)}{\pi(\bold{X}^*)}\\
    &\;\;\;\;+\frac{E\bigg( {0\times \phi_{F^{(1)}}(Y^{(0)})} \mid \bold{X}^*, A=0 \bigg)P(A=0\mid\bold{X}^*)}{\pi(\bold{X}^*)}\Bigg\}\\
    &=E\Bigg\{\frac{E\bigg( {\phi_{F^{(1)}}(Y^{(1)})} \mid \bold{X}^* ,A=1\bigg)\pi(\bold{X}^*)}{\pi(\bold{X}^*)}\Bigg\}\\
    &= E\Bigg\{E\bigg( {\phi_{F^{(1)}}(Y^{(1)})} \mid \bold{X}^* ,A=1\bigg)\Bigg\}\\
    &=E\Bigg\{E\bigg( {\phi_{F^{(1)}}(Y^{(1)})} \mid \bold{X}^* \bigg)\Bigg\}\\
    &=E\bigg\{ {\phi_{F^{(1)}}(Y^{(1)})} \bigg\}\\
    &=0,
\end{aligned}
$$
where the last equality utilizes the property of influence curves (Hampel, 1974). Similar derivations yield that
$$
\begin{aligned}
    \kappa_0 &=E\bigg\{\frac{(1-A)\phi_{F^{(0)}}(Y)}{1-\pi(\bold{X}^*)}\bigg\}\\
    &=0.
\end{aligned}
$$ 
Consequently, combining (\ref{F1F0-expansion}) with (\ref{delat-method-kappa}) gives that, as $n\to\infty$,
$$
     \sqrt{n}\Bigg[\bigg(T(\widehat{F}^{(1)})-T({F}^{(0)})\bigg)-\bigg(T({F}^{(1)})-T({F}^{(0)})\bigg)\Bigg]\overset{d}{\longrightarrow}\text{N}\bigg(0, -\mathcal{B}\mathcal{A}^{-1}\mathcal{B}^\top +\mathcal{C}\bigg),
$$
which gives the result (ii).
\section*{{References}}

\refmark Casella, G. and Berger, R. L. (2002). {\em Statistical Inference, Second Edition}. Duxbury Pacific
Grove, CA.

\refmark Chen, L.-P. and Yi, G. Y. (2021). Analysis of noisy survival data with graphical proportional hazards measurement error model. \textit{Biometrics}, 77, 956-96.

\refmark{Fernholz, L. T. (1983). \textit{Von Mises Calculus for Statistical Functionals}. Springer Science \& Business Media.}

\refmark{Hampel, F. R. (1974). The influence curve and its role in robust estimation. \textit{Journal of the American Statistical Association}, 69(346), 383-393.}

\refmark{Stefanski, L. A. and Boos, D. D. (2002). The calculus of M-estimation. \textit{The American Statistician,} 56, 29-38.}

\refmark{Serfling, R.J. (1980).\textit{ Approximation Theorems of Mathematical Statistics}. John Wiley \& Sons, New York.}

\refmark {Hastie, T., Tibshirani, R. and Wainwright, M. (2015). \textit{Statistical Learning with Sparsity: The Lasso and Generalizations}. CRC press.}

\refmark{Lunceford, J. K. and Davidian, M. (2004). Stratification and weighting via the propensity score in estimation of causal treatment effects: a comparative study. \textit{Statistics in Medicine}, 23, 2937-2960}

\refmark {van der Vaart, A.W. and  Wellner, J.A. (1996).\textit{ Weak Convergence and Empirical
Processes: With Application to Statistics}. Springer-Verlag, New York.}

\refmark{Yang, E., Ravikumar, P., Allen, G. I. and Liu, Z. (2015). Graphical models via univariate exponential family distributions. {\em The Journal of Machine Learning Research}, 16, 3813-3847.}

\end{document}